\documentclass[fleqn,10pt]{wlscirep}
\usepackage[utf8]{inputenc}
\usepackage[T1]{fontenc}
\usepackage{xcolor}
\title{Ubiquitous proximity to a critical state for collective neural activity
in the CA1 region of freely moving mice}

\author[1,2]{Yi-Ling Chen}
\author[1,3]{Chun-Chung Chen}
\author[4]{Yu-Ying Mei}
\author[5]{Ning Zhou}
\author[4]{Dongchuan Wu}

\author[1,6,*]{Ting-Kuo Lee}
\affil[1]{Institute of Physics, Academia Sinica, Taipei ,11529, Taiwan }
\affil[2]{Brain Institute, National Tsing Hua University, Hsinchu City,300, Taiwan }
\affil[3]{Institute of Neuroscience, National Yang Ming University, Taipei,112, Taiwan }
\affil[4]{Graduate Institute of Biomedical Science, China Medical University,Taichung, 404, Taiwan }
\affil[5]{iHuman Institute, ShanghaiTech University ,Shanghai, 201210, China}
\affil[6]{Department of Physics, National Sun Yat-sen University, Kaohsiung, 804, Taiwan}

\affil[*]{tklee@phys.sinica.edu.tw}

\begin{abstract}
Using miniscope recordings of calcium fluorescence signals in the
CA1 region of the hippocampus of mice, we monitor the neural activity
of hippocampal regions while the animals are freely moving in an open
chamber. Using a data-driven statistical modeling approach, the statistical
properties of the recorded data are mapped to spin-glass models with
pairwise interactions. Considering the parameter space of the model,
the observed system is generally near a critical state between two
distinct phases. The close proximity to the criticality is found to
be robust against different ways of sampling and segmentation of the
measured data. By independently altering the coupling distribution
and the network structure of the statistical model, the network structures
are found to be vital to maintain the proximity to the critical state.
We further find the observed assignment of the coupling strengths
makes the net coupling at each site more balanced with slight variation,
which likely helps the maintenance of the critical state. Network
analysis on the connectivity obtained by thresholding the coupling
strengths find the connectivity of the networks to be well described
by a random network model. These results are consistent across different
experiments, sampling and segmentation choices in our analysis.

\end{abstract}
\begin{document}

\flushbottom
\maketitle

\section*{Introduction}
The CA1 region of hippocampus is an area of brain that is important
for representing space information of the environment \cite{okeefe1976placeunits}.
Notably, place cells, the neurons that respond to a certain location
or landmark in the environment, are first discovered in this region.
Recently developed technologies such as miniscopes \cite{ghosh2011miniaturized}
allow optical recording of large numbers of neurons in this region while the animal is freely behaving
in an experimental environment. Through calcium ($Ca^{2+}$)-dependent fluorescence,
the firing activities of neurons can be inferred simultaneously. This
opens up the possibility for studying population coding in these regions
of the brain.

To understand implications of the large volume of data, maximum-entropy
modeling \cite{schneidman2006weakpairwise,tkacik2006isingmodels}
has been used to match the statistical properties of the observed
firing images to that of a pairwise-interaction model of binary spin
glass. While only matching the first and second order correlation
statistics of the spins, such models have been shown to well reproduce
the higher-order correlations as well as other properties of the observed
data for the neural systems \cite{meshulam2017collective,tkavcik2015thermodynamics}.
Generalizing the model system to a broader parameter space, it has
also been shown that the original parameter values corresponding to
the observed brain images are poised near a critical point in the
parameter space, consistent with the so-called critical brain hypothesis
\cite{mora2011arebiological,usher1995dynamic,beggs2008thecriticality}.

To find the best values of model parameters that reproduce the statistical
properties of observed data, we use the Boltzmann learning (BL) method
\cite{ackley1985alearning}, which amounts to performing gradient
descent on the Kullback--Leibler (KL) divergence \cite{kullback1951oninformation}
between the observed and modeled distributions of the binarized system
states.

For many of the reported cases in statistical modeling of neural dynamics,
in the proximity of a critical state seems to be a common observation
for these systems. Here we refer the system to be in the proximity
of a critical state simply because the true critical state should
be considered for an infinitely large system while we only consider
60 to 100 neurons here. Even when the system size is finite, many
properties are still quite similar to the critical state \cite{stanley1987introduction}.
Since the recorded neurons are only a very small fraction of the total
neurons, it is desirable to verify whether the same conclusions would
have been reached had the systems been sampled or segmented differently.
Based on the miniscope data from four different experiments, the fitted
models corroborate the robustness of the proximity to a critical state
under different subsampling and segmentation processes. The recovered
best-fitted parameters of the external fields and the couplings in
statistical modeling can reproduce the observed mean and pairwise
correlations accurately. Similar distributions of these structural
parameters are found under different sampling and segmentation conditions.

Having established the stability of the model parameters, we further
analyze the connectivity of the inferred spin networks with different
thresholds on the coupling strength and calculate their network properties,
such as, cluster coefficient, average path length, and degree distribution.
The network turned out to be well-described by a random network model
\cite{erdhos1959onrandom} for all experiments and subsamples when
various thresholding criteria of connectivity are applied.

    The main result is presented
in section Results. The structure of the coupling strength
parameters and the criticality of the systems are examined in section Parameters of spin-glass model. 
In section Analysis of network connectivity.
the network connectivity is analyzed with different thresholds. It
shows the property of a random network model. The discussion is in
the section Discussion. The statistical modeling is described in section Methods.

\section{Results \label{sec:results}}

\begin{figure}[!h]
\begin{centering}
\includegraphics[width=12cm]{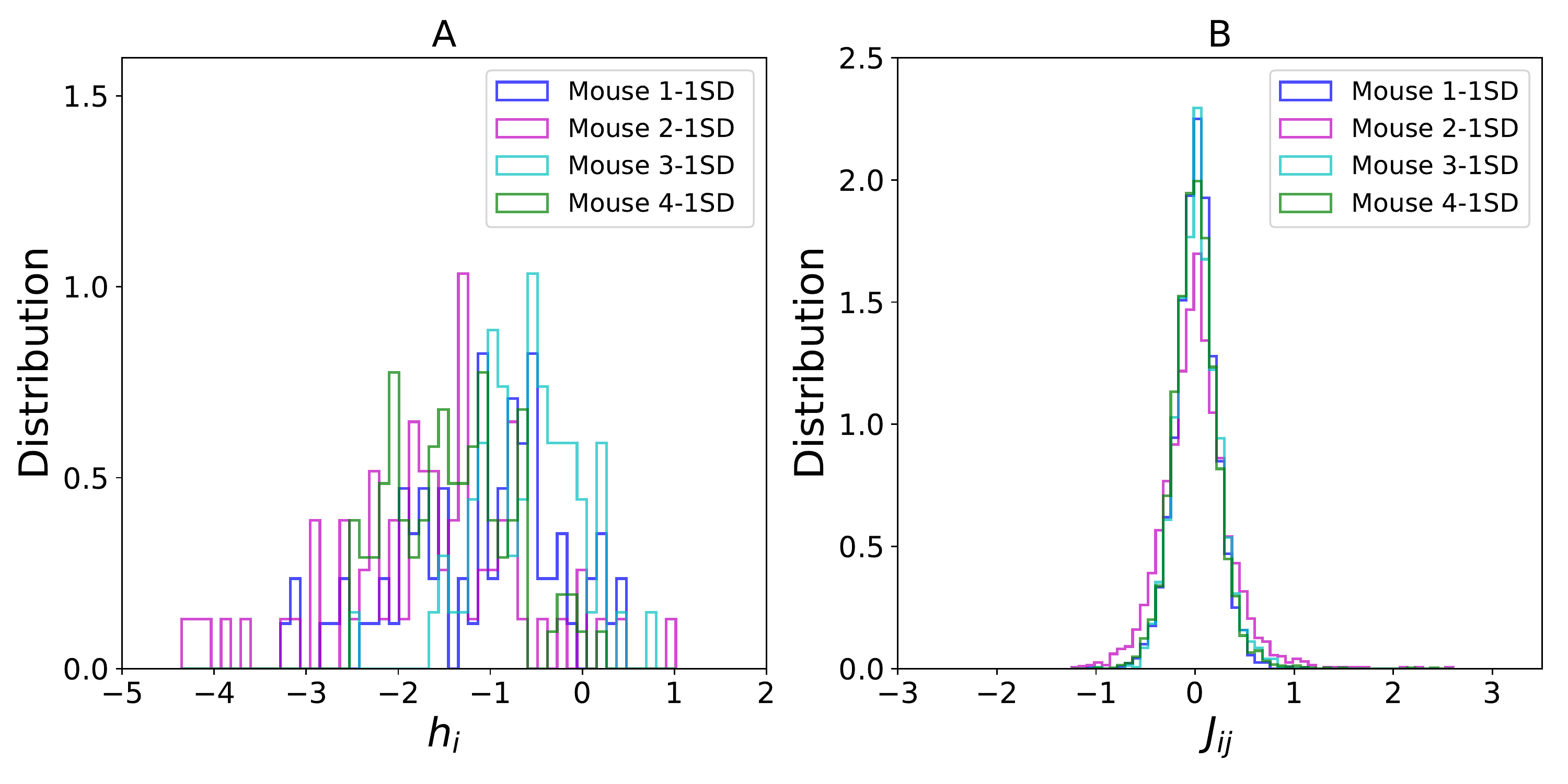}
\par\end{centering}
\caption{{\bf Distributions of best-fitted local field and coupling.}
\textbf{A.} The distributions of the best-fitted local field $h$
in the four mice \textbf{B.} The distributions of the best-fitted
coupling $J$.}
\label{fig:allsystem_MouseJ}
\end{figure}

In order to compare the fitted parameters of local field $h$ and
coupling $J$ for the four datasets, their distributions are calculated
and shown in Fig.~\ref{fig:allsystem_MouseJ}A and Fig.~\ref{fig:allsystem_MouseJ}B,
respectively. The detail is described in the Supplementary S1. The mean values and
their standard deviations are listed in Table~\ref{tab:h-J}. Since
neurons or the identified regions of interests (ROIs) in these systems in general favor silent states, the mean values
of the local fields are all negative. The distributions of $h$ are
quite broad but mostly have values within the same range of -3 to
1 for the four cases. The mean values of $J$ are all very small and
it is close to the Gaussian distribution except the long tails at both
ends. Actually, the long tail at the positive end is a bit more extended
than the Gaussian. More detailed validation of the parameters is in Supplementary S2 
and model predictions in the Supplementary S3.

\begin{table}[!ht]
\centering
\caption{
{\bf The means and standard deviations (in parenthesis) of the distributions
shown in Fig.~\ref{fig:allsystem_MouseJ}.}}
\begin{tabular}{l|l|l}
\hline
{\bf dataset} & {\bf local field $h$} & {\bf coupling $J$}\\ 
Mouse 1-1SD & $-1.12(\pm0.87)$ & $0.01(\pm0.21)$ \\ \hline
Mouse 2-1SD & $-1.68(\pm1.04)$ & $0.01(\pm0.33)$\\ \hline
Mouse 3-1SD & $-0.60(\pm0.55)$ & $0.02(\pm0.22)$\\ \hline
Mouse 4-1SD  & $-1.41(\pm0.63)$ & $0.01(\pm0.23)$\\ \hline
\end{tabular}
\label{tab:h-J}
\end{table}

\subsection{Robustness of the critical behavior of the four datasets \label{sec:Robustness-of-the}}

\begin{figure}[!h]
\begin{centering}
\includegraphics[width=10cm]{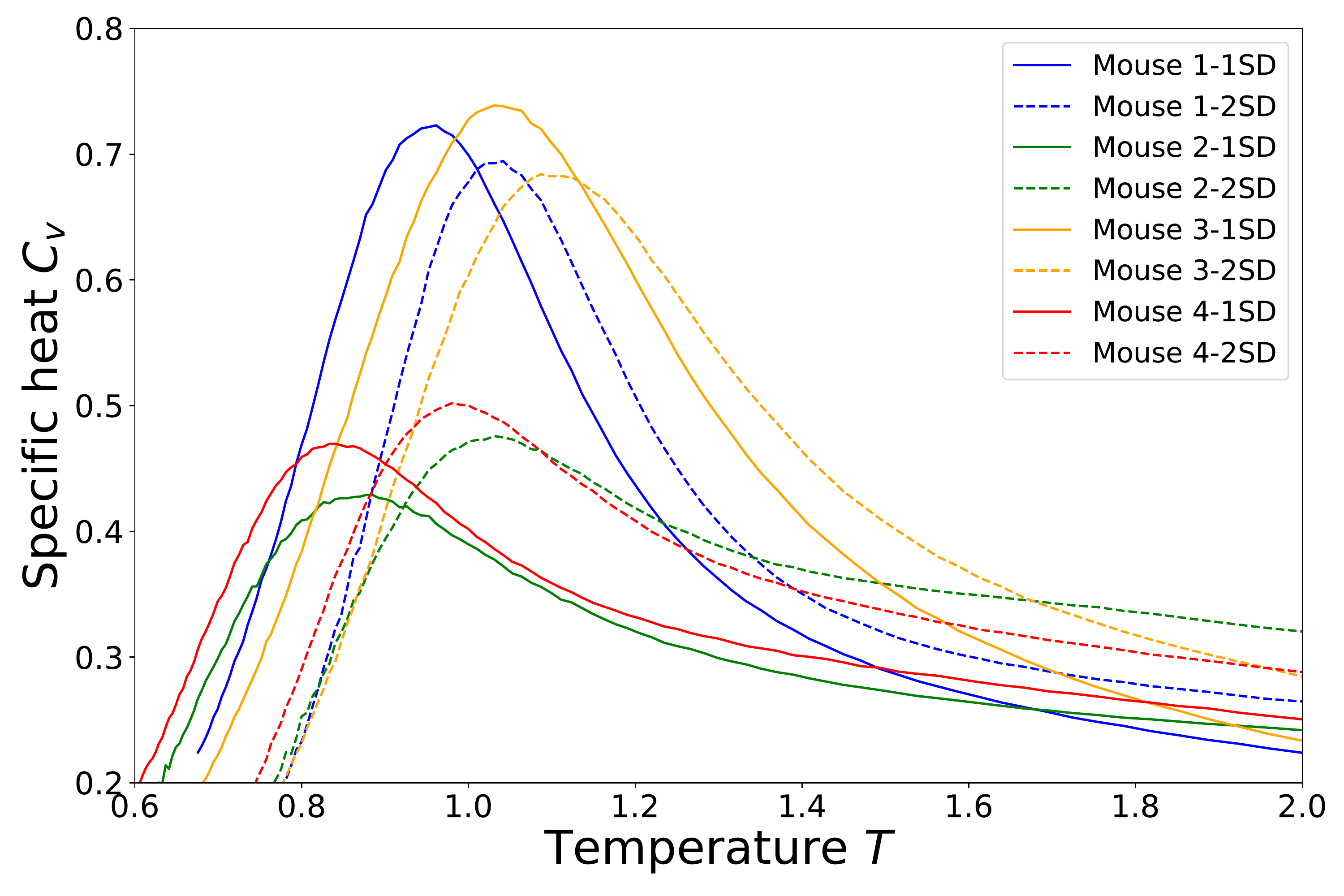}
\par\end{centering}
\caption{{\bf Specific-heat curves.}
Specific-heat curves for statistical models for the four experiments.
The experimental data is at $T=1.0$. The solid (dashed) curves are
obtained with segmentation threshold of 1 and 2 standard deviation
(SD) $\sigma\left(L_{i}^{\text{Ca}^{2+}}\right)$ for the calcium-dependent
fluorescence trace of each ROI $i$.}
\label{fig:Specific-heat-curves}
\end{figure}

To characterize the thermodynamic properties
of the model, we extend the parameter space to include arbitrary
$T$ and calculate the specific heat of the model 
\[
C_{v}=\frac{1}{NT^{2}}\sigma^{2}\left(E\right)
\]
where $\sigma^{2}\left(E\right)$ is the variance of energy, as a
function of temperature. As commonly seen in earlier studies \cite{mora2011arebiological,meshulam2017collective},
the specific heat curve exhibits a single peak around the original
system temperature $T=1$ for all four datasets from the experiments (Fig.~\ref{fig:Specific-heat-curves}). 

In the thermodynamic limit where the system has an infinite number
of spins \cite{landau1976finitesize}, the specific-heat peak can
diverge and signify a critical point of the system. However, as shown
in Ref~\cite{landau1976finitesize}, when there is only a finite
number of spins, the specific heat will have a peak near the critical
temperature. The peak becomes broader and peak temperature ($T_{\text{p}}$)
moves away from the true critical temperature for systems with fewer
and fewer spins. Our result shows $T_{\text{p}}$ close to 1, where 
$T=1$ is the temperature model parameters are determined from the dataset. 
This indicates that the observed system has a thermodynamic temperature near the critical
point of our deduced spin-glass model. In other words, the observed
neural state happens to be close to the boundary of two phases in
the model spin system. It should be noticed that the thermodynamic
temperature mentioned here is meaningful only in the model-parameter
space and it is not related to the physical temperature of the real
neural system. In our discussion below, we will identify the $T_{\text{p}}$
of specific heat as the critical temperature.

\begin{figure}
\begin{centering}
\includegraphics[width=10cm]{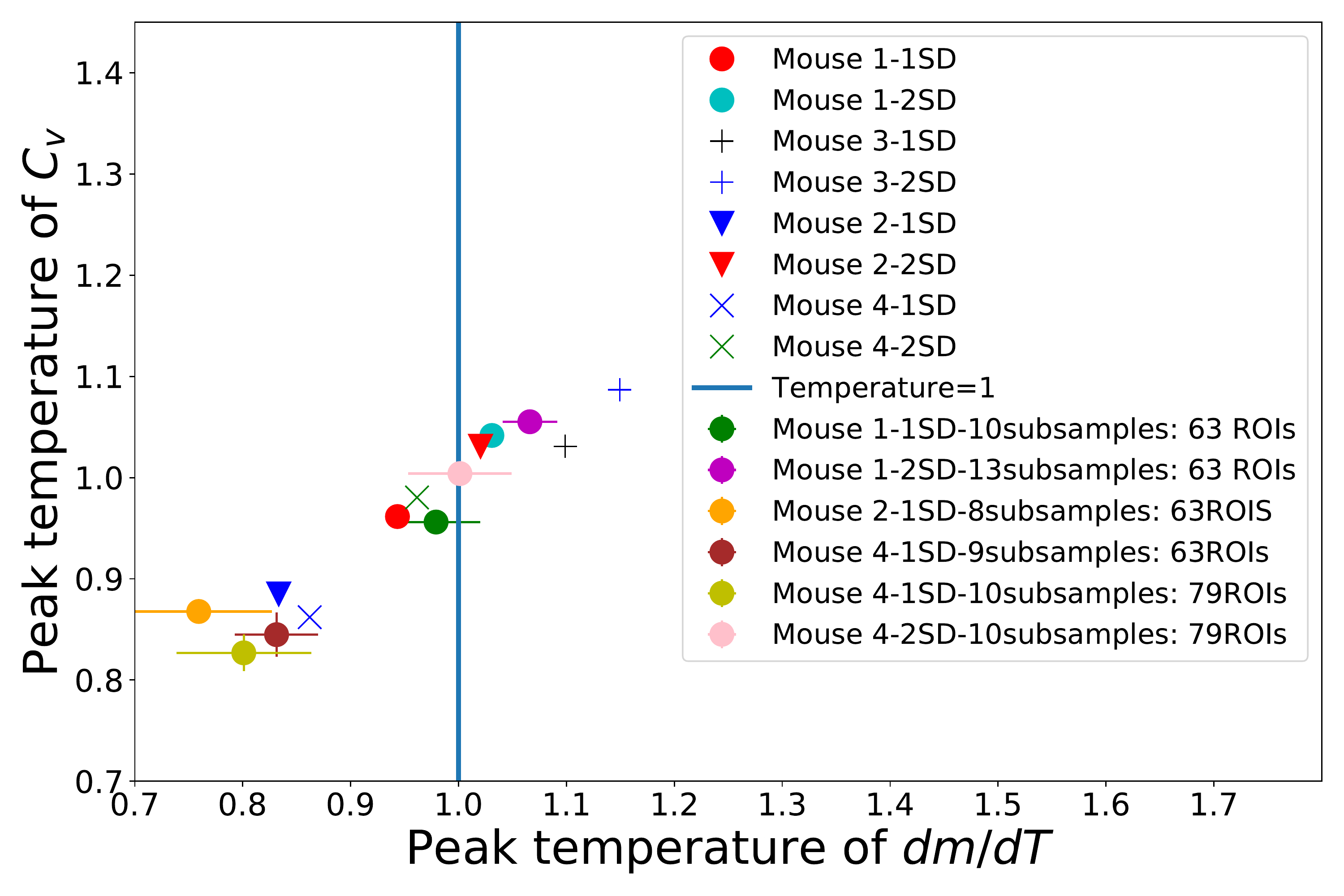}
\par\end{centering}
\caption{{\bf Peak temperature of $dm/dT$ versus Peak temperature of $C_{v}$.}
Critical temperature determined by the peak of specific heat (vertical
axis) versus that determined by the peak of magnetization slope (horizontal
axis) for the four mice under different ways of segmentation and subsampling.}
\label{fig:Critical-temperature-determined}
\end{figure}

To make sure that the observed critical state is not incidental to
our choice of segmentation threshold, it is important to repeat the
calculations using 2 times the standard deviation of each calcium
trace as the binarization threshold. The resulting specific-heat curves
using both $\sigma\left([\text{Ca}^{2+}]_{i}\right)$ (referred to as 1SD from now on) and 2$\sigma\left([\text{Ca}^{2+}]_{i}\right)$
(2SD) as thresholds are shown in Fig.~\ref{fig:Specific-heat-curves}.
The peak positions of specific heat as listed in Table~\ref{tab:Peak-temperature}
are all close to $T=1$. In addition to the peak temperature of the
specific heat, the critical point of the spin system can also be identified
with the maximum of the slope $dm/dT$ \cite{huang1988statistical}, where
the magnetization $m$ is thermodynamic average of the sum of all
the spins. The detail about $m$ vs $T$ is discussed in the Supplementary S3 temperature dependence of the magnetization
. To compare
the critical temperatures obtained from both criteria, the temperatures
of peak magnetization slope $dm/dT$ are plotted against the temperatures
of peak specific heat $C_{v}$ for all calculated cases of various
thresholds, subsamples, and mice in Fig.~\ref{fig:Critical-temperature-determined}.
The plot of magnetization slope $dm/dT$ vs. temperature for Mouse 1-1SD 
is shown in the Supplementary Figure S7.
This is the most important result of this work that we find these
critical temperatures within $20\%$ of $T=1$ showing a ubiquitous
proximity to a critical state as listed in Table~\ref{tab:Peak-temperature}
for all cases. The data in Fig.~\ref{fig:Critical-temperature-determined}
also reveals an interesting correlation that the peak temperatures
of both $dm/dT$ and $C_{v}$ are either greater or smaller than 1
together. It is well known for this kind of model \cite{stanley1987introduction}
with a finite number of spins, the peaks of $dm/dT$ and $C_{v}$
will move away from the ideal critical temperature in the same way.

\begin{table}[!ht]
\centering
\caption{
{\bf Peak temperature of $dm/dT$ versus peak temperature of $C_{v}$ for
four different experiments.} Using different thresholds for segmentation,
and some random subsamples of given size from the measured data. For
multiple subsamples of the same size and from the same data, the mean
values of the peak temperatures are shown followed by their standard
deviations in parenthesis.}
\begin{tabular}{l|l|l}
\hline
{\bf dataset } & {\bf peak of $dm/dT$} & {\bf peak of $C_{v}\left(T\right)$}\\ 
Mouse 1-1SD & 0.94 & 0.96 \\ \hline
Mouse 1-1SD-10subsamples: 63 ROIs & $0.98(\pm0.04)$ & $0.96(\pm0.01)$\\ \hline 
Mouse 1-2SD& 1.03 & 1.04\\ \hline 
Mouse 1-2SD-13subsamples: 63 ROIs & $1.06(\pm0.03)$ & $1.05(\pm0.01)$\\ \hline 
Mouse 2-1SD& 0.83 & 0.88\\ \hline
Mouse 2-1SD-8subsamples: 63 ROIs & $0.76(\pm0.07)$ & $0.87(\pm0.01)$\\ \hline 
Mouse 2-2SD& 1.02 & 1.03\\ \hline
Mouse 3-1SD& 1.10 & 1.03\\ \hline
Mouse 3-2SD& 1.15 & 1.09\\ \hline 
Mouse 4-1SD& 0.81 & 0.84\\ \hline
Mouse 4-1SD-9subsamples: 63 ROIs & $0.83(\pm0.04)$ & $0.84(\pm0.02)$\\ \hline 
Mouse 4-1SD-10subsamples: 79 ROIs & $0.8(\pm0.06)$ & $0.83(\pm0.02)$\\ \hline
Mouse 4-2SD& 0.96 & 0.98\\ \hline
Mouse 4-2SD-10subsamples: 79 ROIs& $1(\pm0.05)$ & $1(\pm0.01)$\\ \hline
\end{tabular}
\label{tab:Peak-temperature}
\end{table}

\subsection{Robustness of the critical behavior of the random subsamples}

\begin{figure}
\begin{centering}
\includegraphics[width=10cm]{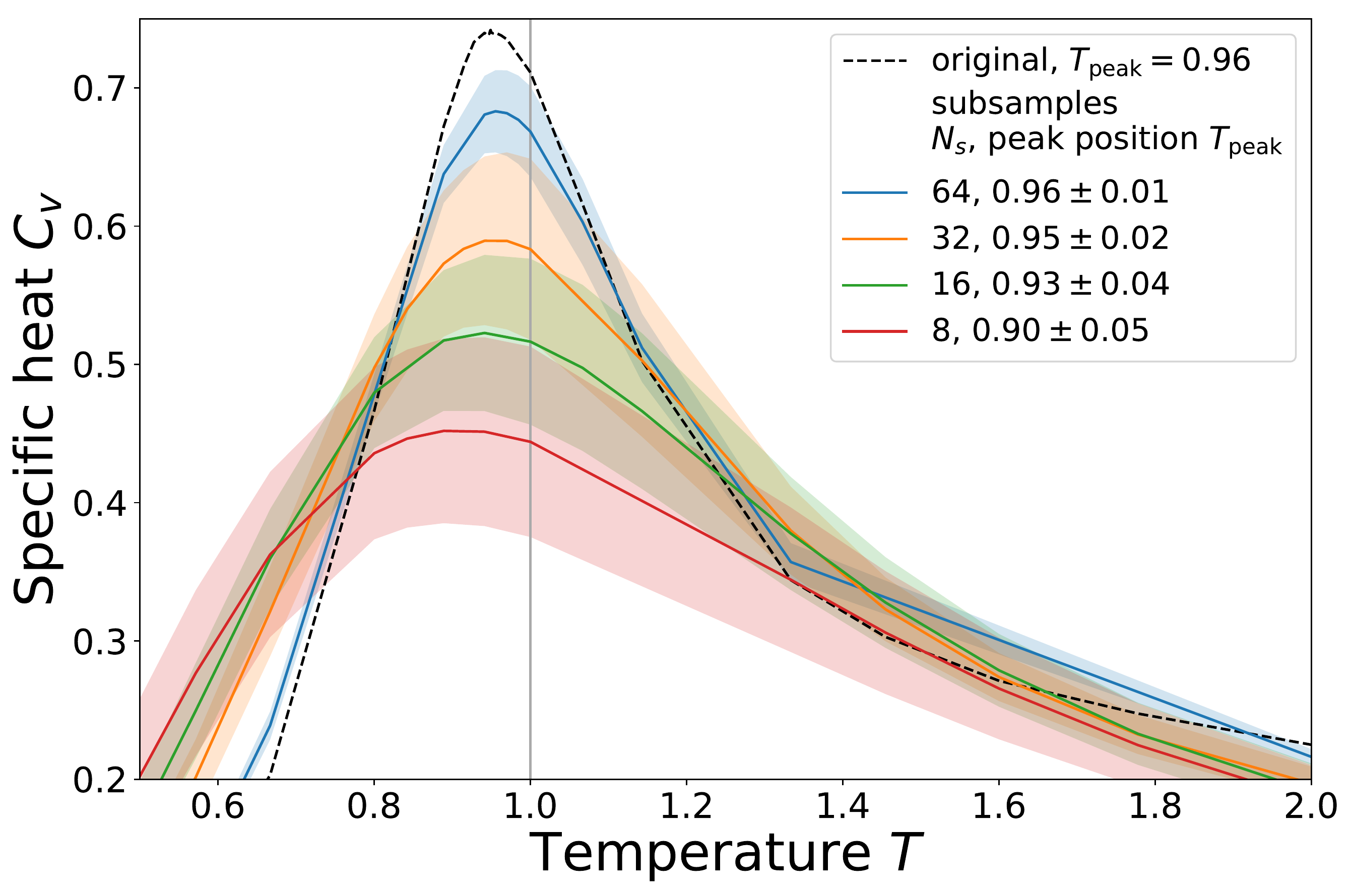}
\par\end{centering}
\caption{{\bf Specific heat curve of original Mouse 1-1SD and subsamples.}
Specific heat as a function of temperature for random subsamples of
measured ROIs in Mouse 1-1SD experiment. The shaded regions represent the spread of
the results for 16 samples for a given size of subsamples.}
\label{fig:Specific-heat-subsample}
\end{figure}

While the volume of the acquired data in the experiments may be large,
it typically represents only a very small portion of the neural systems.
Or, the measurements may not have single-cell resolution. Additionally,
the way of segmenting the recordings into discrete states may further
discard a significant amount of the information from the measurements.
With all these uncertainties, it requires us to further examine the
conclusion of the last section that the data represents a state in
the proximity of a critical state. One approach to address this issue
is by studying subsamples consists of a subset of ROIs randomly selected from 
the original experimental set. Then, each is treated as a new set of data and 
the calculations are repeated to find the best fitted model. Here we consider random
subsets of sizes $N_{s}$ = 64, 32, 16, and 8. Each size of the subsamples is repeated 16 times
to estimate the standard deviation for the specific heat and the peak
position as shown in Fig.~\ref{fig:Specific-heat-subsample}.

The resulting peak positions of the specific heat show a weak dependence
on the system size within 10\% down to $N_{s}=8$. Similar with the
result discussed by \cite{landau1976finitesize} for a system of
nearest neighbor Ising model, the peak becomes broader and peak temperature
further shifts away as the number of spins is reduced. Here we only
show the subsamples of one dataset, Mouse 1 with 1SD. Similar result
from the Mouse 4 with 1 and 2 SD is shown in the Supplementary Figure S8.

\subsection{Parameters of spin-glass model\label{sec:Structural-parameters-of}}

The coupling parameters of the spin-glass model obtained from statistical
modeling constitute an effective or functional network structure that
can reproduce the observed state properties. One can expect that such
network structure may bear some significance pertinent to the functional
dynamics of the brain. However, before such possibility is pursued,
we need also to make sure the structure is quite generic with little
dependence on segmentation and sampling. For all of results reported
above, the coupling strength $J$ has much larger influence than the
local field $h$ on the statistical properties of the model as long
as $h$ is within the range shown in Fig.~\ref{fig:allsystem_MouseJ}A.
Hence, only coupling $J$ will be considered in the analysis below.

\subsubsection{Distribution and values of coupling strengths}

\begin{figure}
\begin{centering}
\includegraphics[width=10cm]{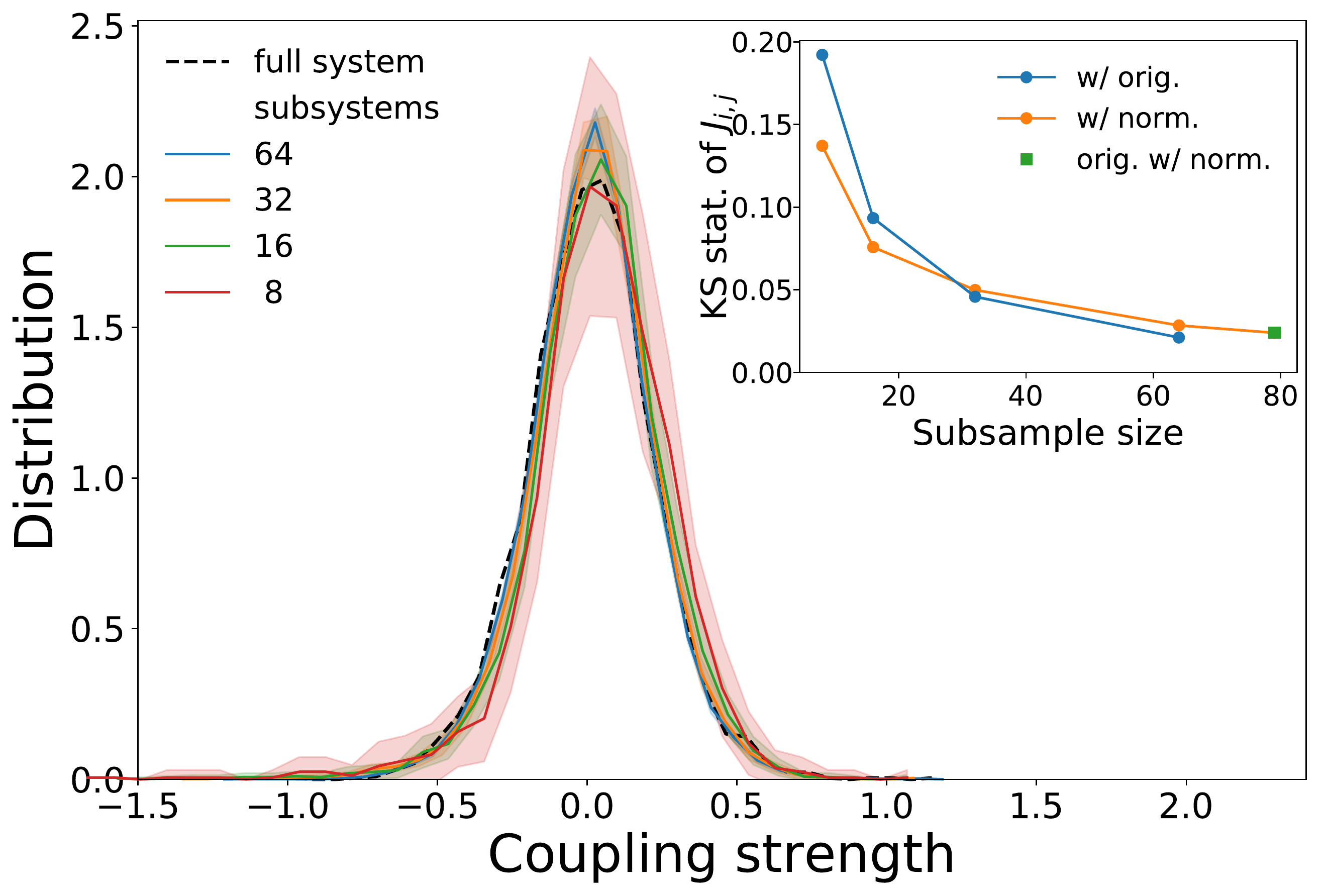}
\par\end{centering}
\caption{{\bf Distributions of coupling strength of original Mouse 1-1SD and subsamples.}
Distributions of coupling parameter $J$ of the statistical models
for the subsampled systems (drawn lines) compared with that for the
original system (dashed line). The shaded area is the standard deviation
estimated using 16 randomly drawn subsamples for each subsample size.
The inset shows the Kolmogorov--Smirnov (KS) statistics of the subsamples
with original (orig.) and a Gaussian (norm.) of the same mean and variance.
The KS statistic between the distribution for the original 79 ROIs
of the Mouse 1-1SD system and a Gaussian distribution of the
same mean and variance is marked with a square.}
\label{fig:Distributions-of-coupling}
\end{figure}

To find out whether the distributions of coupling strength $J$ obtained
in Fig.~\ref{fig:allsystem_MouseJ}B is sensitive to the number of
neurons detected, subsets of neurons were randomly selected to form
a number of subsamples. Then the model parameters of each subsample
were calculated by the BL method. In Fig.~\ref{fig:Distributions-of-coupling},
the coupling-strength distribution of the subsystems is compared to
that from the full dataset and quantified using the Kolmogorov--Smirnov
(KS) test (see the Supplementary S4 Kolmogorov–Smirnov test for the definition of KS statistic)
. Similar with the
temperature dependence of specific heat, the KS statistic of subsamples
shows weak size dependence and remains quite small until the subsample size reaches below 30 as shown in the
inset of Fig.~\ref{fig:Distributions-of-coupling}.

 The robustness of the coupling strength distribution under the subsampling
gives us an encouraging hope that, in our approach, it may not be
necessary to measure all the neurons but only a relatively small subset
of neurons to obtain a similar result.

\begin{figure}
\begin{centering}
\includegraphics[width=10cm]{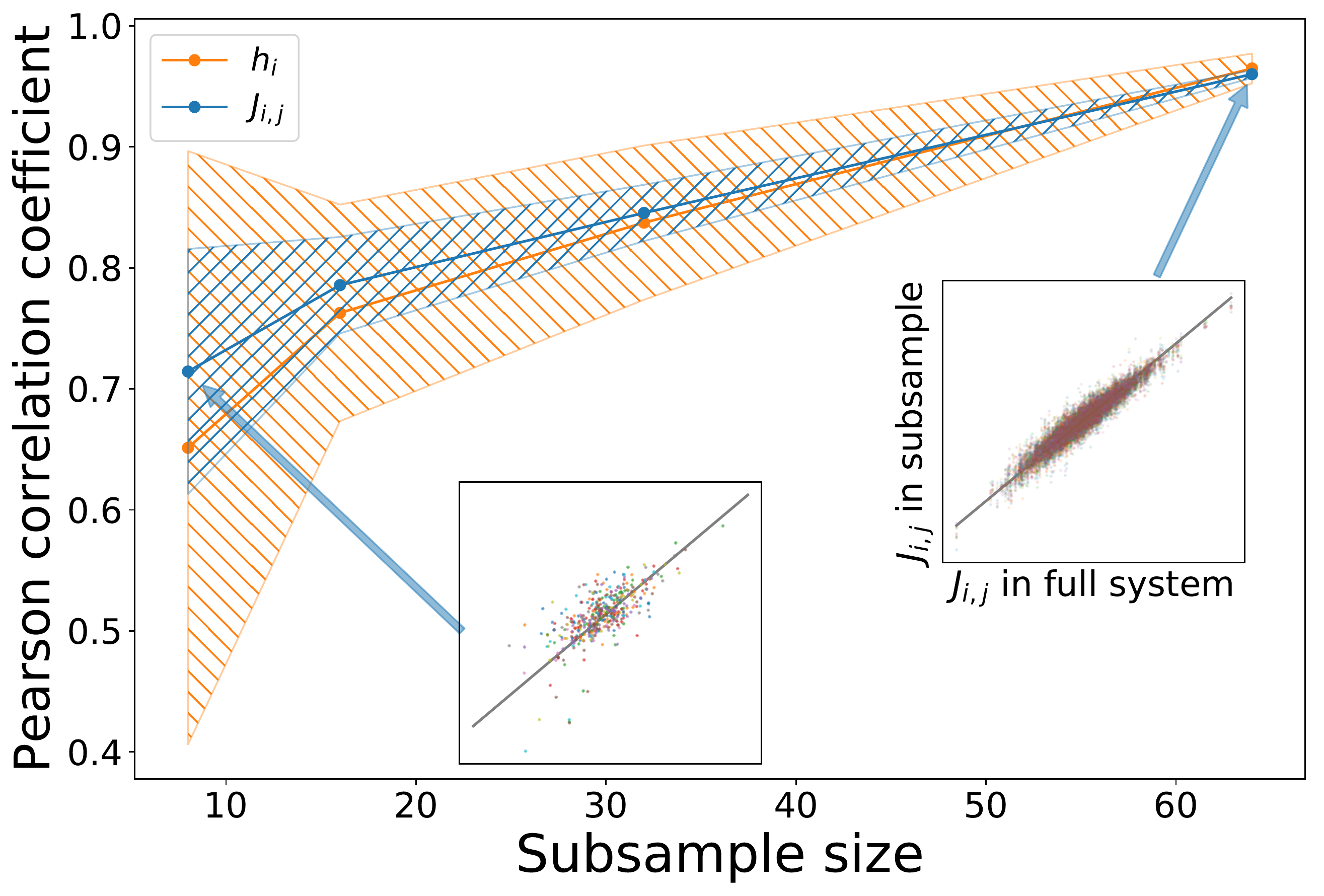}
\par\end{centering}
\caption{{\bf Pearson correlation coefficients of the best-fitted local field and coupling between the original Mouse 1-1SD and subsamples. }
Pearson correlation coefficients of the parameters $J_{ij}$ and $h_{i}$
between the models of subsamples and the model for the original system
of Mouse 1-1SD. The coefficients remain above 0.6 down
to subsample size 8. The insets show comparisons of $J_{ij}$ values
between the subsamples (vertical axis) and the original (horizontal
axis).}
\label{fig:Pearson-correlation-coefficients}
\end{figure}

Besides overall distribution, the value of the coupling strength
between any two spins in a subsystem is compared to the coupling strength
between the same spins in the original full system. Since, in principle,
two very different datasets are separately fitted, the Pearson correlation
coefficient between them should be very small. However, Fig.~\ref{fig:Pearson-correlation-coefficients}
shows the coefficients for the different subsamples all above 0.6.
It is important to note that the subsystems have a very strong correlation
with the original much larger system with Pearson coefficients between
0.75 to 0.9 for a system with only $40\%$ neurons left. Details about
calculations of Pearson correlation coefficients can be found in the Supplementary S4 Pearson correlation coefficients.
Then, the next question is whether having a very similar pair-coupling
structure is what allows the subsample systems to remain in the
proximity of the critical state as shown in Figs.~\ref{fig:Critical-temperature-determined}
and ~\ref{fig:Specific-heat-subsample}.

\subsubsection{Importance of network structure}

\begin{figure}
\begin{centering}
\includegraphics[width=10cm]{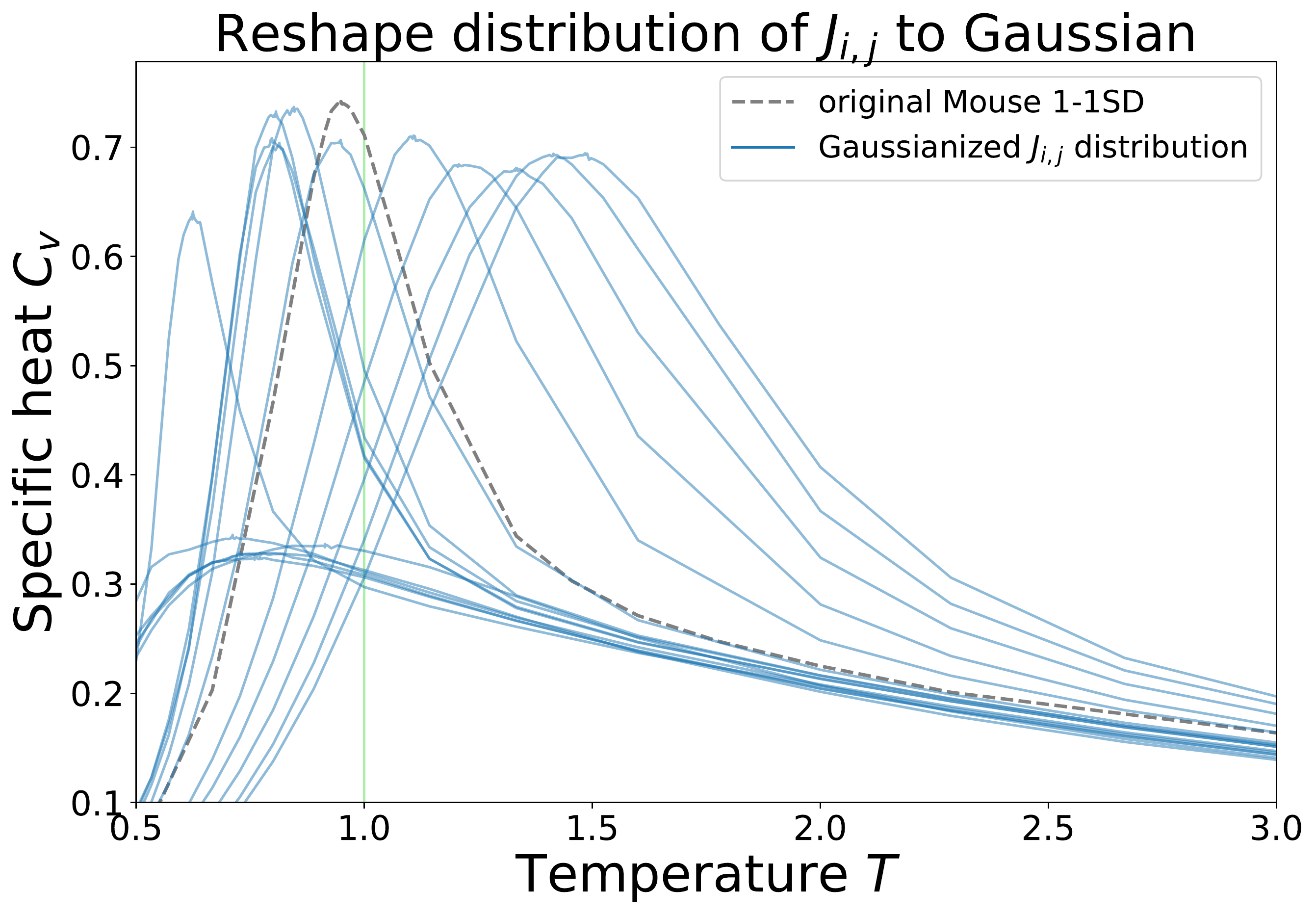}
\par\end{centering}
\caption{{\bf Specific-heat curves of the original Mouse 1-1SD and Gaussians. }
Specific-heat curves of models with new coupling strength values randomly
drawn from a Gaussian distribution of the same mean and variance as
the coupling distribution of the original model. The rank of coupling-strength
order of the spin pairs are preserved when these new strength values
are assigned.}
\label{fig:Reshaped}
\end{figure}

To answer the above question, it is important to first understand the unique
characteristics of the pair-coupling structure. Thus two
alterations to the coupling structure of the model are considered. Firstly, we reshape
the coupling strength distribution into a Gaussian with the same mean
and variance and randomly redraw the values of coupling strength while
maintaining the rank of strength order for all spin pairs. The resulting
specific heat-curves shown in Fig.~\ref{fig:Reshaped} show a
broad variation of peak temperatures. Unlike the original data
shown in dashed line, most subsamples have the peak temperatures of
specific heat far away from $T=1$, which is the state related to
the recorded data.

\begin{figure}
\begin{centering}
\includegraphics[width=10cm]{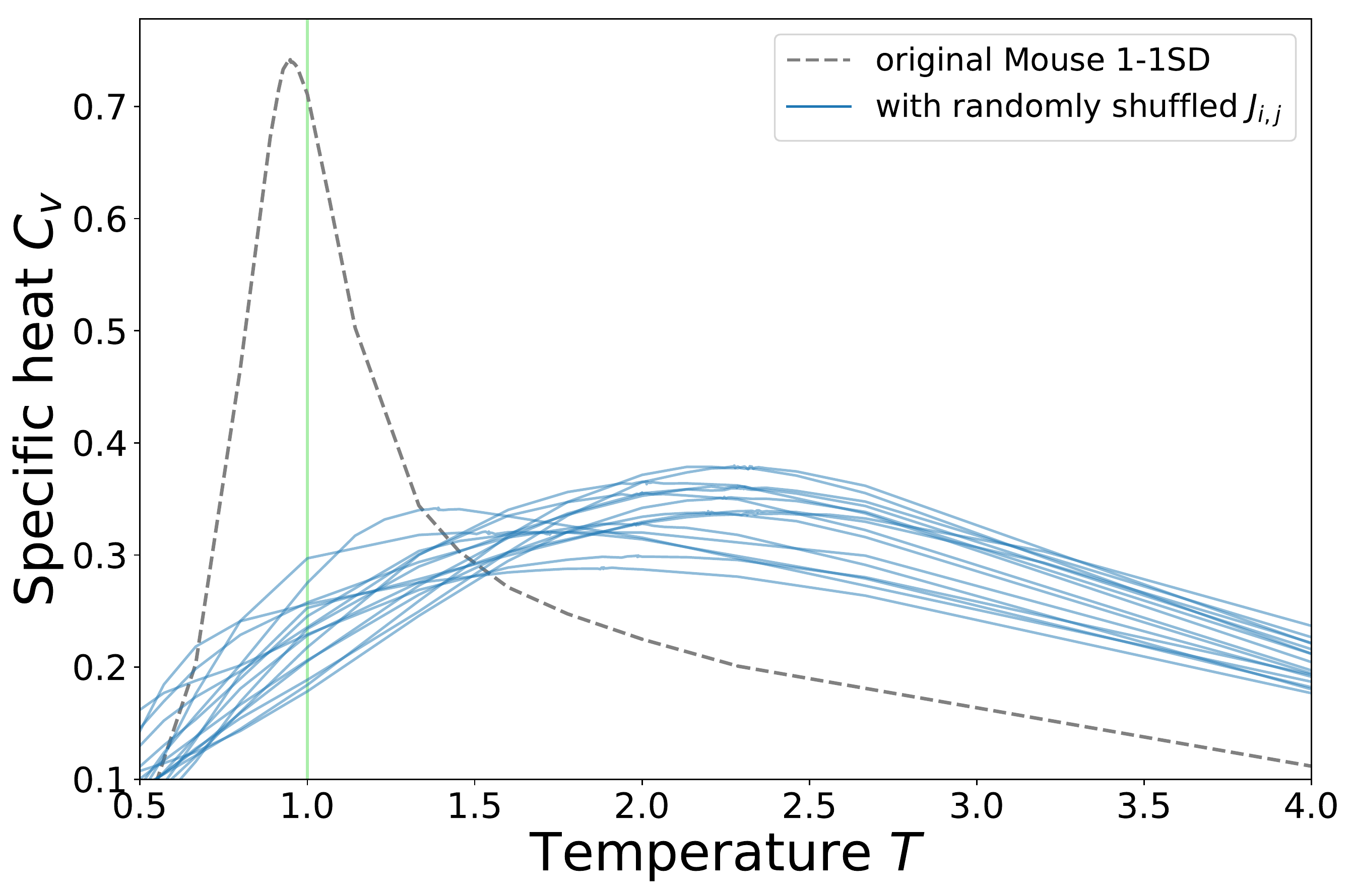}
\par\end{centering}
\caption{{\bf Specific-heat curves of original Mouse 1-1SD and shuffled network structures. }
Specific-heat curves of models with randomly shuffled network structures
from the original statistical model of Mouse 1-1SD.
The manipulation keeps the set of coupling strength values and reassigns
them to different spin-pairs.}
\label{fig:randomized-network}
\end{figure}

In the second case, the pair-coupling strength $J_{ij}$ is randomly
shuffled. This manipulation destroys the network structure while preserving
the distribution of coupling strength. All the resulting specific-heat curves
as shown in Fig.~\ref{fig:randomized-network} have their
peaks shifted to the right with the average value $\bar{T}_{\text{p}}\approx2.05$.
This implies that at $T=1$, systems represented by the shuffled $J$
are in the low temperature phase of their respective models. The two
studies in this subsection reveal a couple of important characteristics
of the pair-coupling structure. Randomly assigned coupling strength
between pairs, in general, will not lead to a pair-interaction model
with the peak of specific heat sitting near $T=1$. The specific
pair-coupling structure obtained from the experiment must be maintained
to keep the state in the proximity of a critical state.

\subsubsection{Balanced network in net coupling}

\begin{figure}
\begin{centering}
\includegraphics[width=10cm]{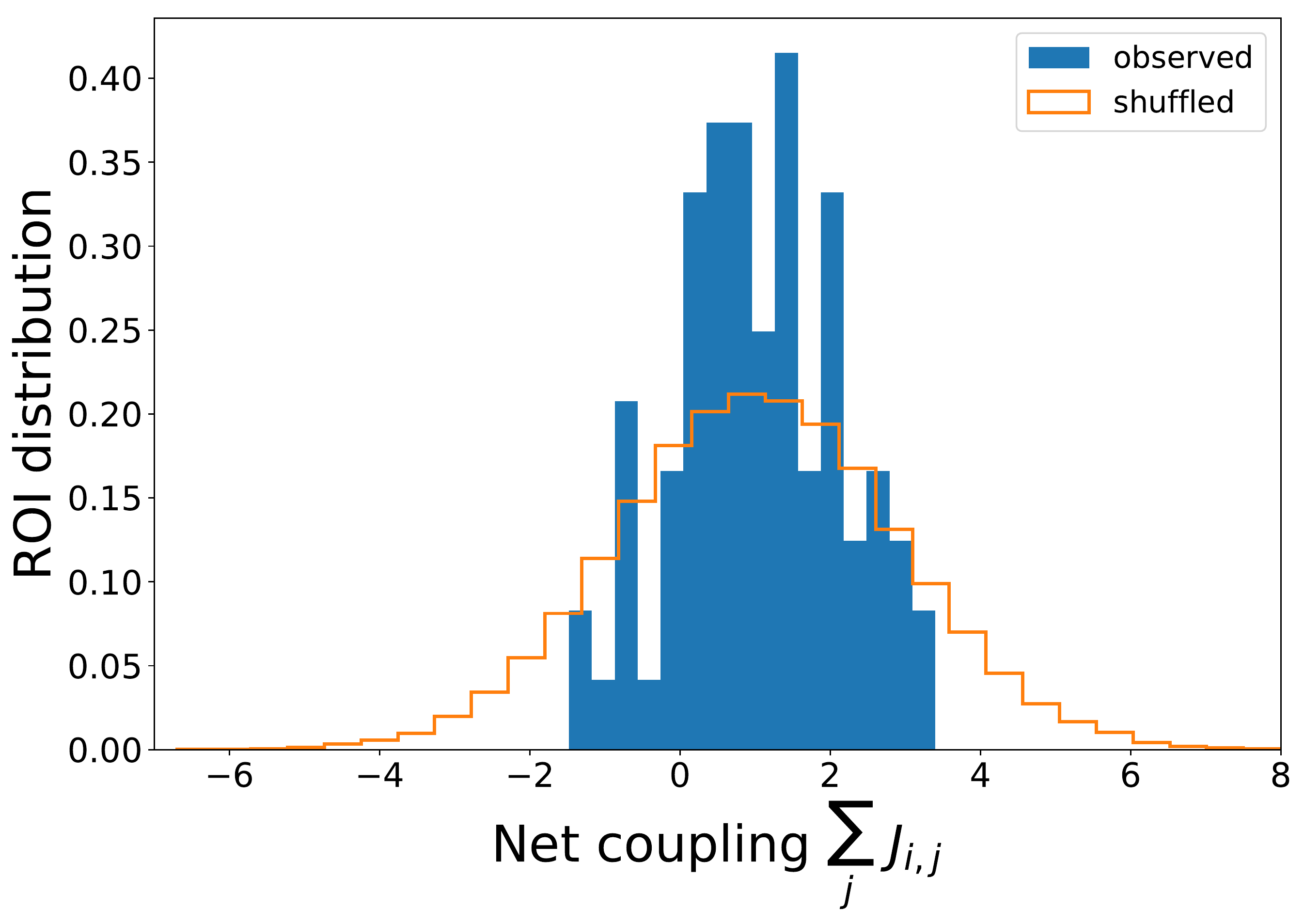}
\par\end{centering}
\caption{{\bf Distribution of net couplings for the ROIs of Mouse 1-1D. }
Distribution of net couplings $J_{i}^{\text{net}}\equiv\sum_{j}J_{ij}$
for the ROIs in the model of Mouse 1-1SD dataset (shaded
area), compared with the average of net ROI coupling distribution
of shuffled networks from the same coupling $J_{ij}$ distribution.}
\label{fig:Distribution-of-net}
\end{figure}

Further investigation of the resulting network reveals that the observed
network is generally more \emph{balanced} compared to a randomly shuffled
model. That is, if we define the net coupling of an ROI as the sum
of the coupling strengths of all links connecting to this ROI, $J_{i}^{\text{net}}=\sum_{j}J_{ij}$,
the distribution of net couplings for the observed network is much
narrower compared to a typical randomly shuffled samples as shown
in Fig.~\ref{fig:Distribution-of-net}.

The narrower distribution of net ROI coupling can be
observed for all our four experiments. (See the Supplementary S4 Distributions of net coupling strength of ROIs.) To see how the balance
of the coupling strength distribution affects the criticality of a
network, we repeatedly apply selective random swapping of the link
strengths of a shuffled network in order to make its net-ROI-coupling
distribution narrower. We find the peak of its specific-heat curve
does move towards $T=1$ as net couplings become more uniform. However,
this process alone is still insufficient to make the model as critical
as what is observed experimentally. For the model of Mouse 1-1SD, the peak temperature of specific heat moves from $2.05\pm0.28$
to $1.65\pm0.21$ when randomly shuffled model is rebalanced to as good
as the observation, and to $1.35\pm0.25$ at the limit of the balancing
update when the net ROI coupling distribution is twice as narrow as
the observation. Again, it is not easy to obtain a model with the peak 
temperature at $T= 1$.

\subsection{Analysis of network connectivity\label{sec:network connectivity}}

In the section discussed above, the network structure obtained from
the mice data shows several very interesting characteristics that
are important to the results presented so far. But what kind of network
structure is it? Here we will try to answer this question. First, let
us define some quantities that are commonly used in classifying network
structure. For a connected network $G$ consists of the set of vertices
(nodes or ROIs) $V$, the average local clustering coefficients
of the network is given by \cite{smallworld,smallworld1} 
\begin{equation}
\mathrm{CC}(G)=\frac{1}{\lvert V\lvert}\sum_{v\in V}\mathrm{CC}(v),
\end{equation}
where the local clustering coefficient $\mathrm{CC}(v)$ of a vertex
$v$ is defined as 
\begin{equation}
\mathrm{CC}(v)\equiv\frac{2N_{v}}{k_{v}(k_{v}-1)},
\end{equation}
where $k_{v}$ is the degree or the number of edges/bonds of $v$, and $N_{v}$ is the number of
connected vertex pairs whose both vertices are connected to $v$.
The path length $d\left(v_{1},v_{2}\right)$ is the minimal number
of links that connect the vertices $v_{1}$ and $v_{2}$ to each other.
The average path length $L(G)$ of a graph $G$ is the average path
lengths over all vertex pairs in the graph \cite{smallworld,smallworld1}.

In the model, the coupling $J_{ij}$ has both positive and negative
values, hence it is separated into excitatory and inhibitory network
with positive and negative $J$, respectively. To examine the structure
more carefully, the properties of excitatory network will be calculated
with a lower threshold $X_{l}$ for $J_{ij}$, so that only bonds
or edges with $J_{ij}>X_{l}$ will be considered. For larger $X_{l}$,
there are few edges or bonds between nodes and few nodes are connected.
But as $X_{l}$ decreases, more connections are forming and all the
nodes may become connected. Thus the degree distribution, clustering
coefficient and path length will all depend on the lower threshold
value $X_{l}$. There is a critical value $X_{l}^{c}$ that for $X_{l}>X_{l}^{c}$,
the network is disconnected such that not all the nodes are connected in one maximum cluster and we have
isolated clusters or nodes. Similarly, properties of inhibitory network changes with the upper
threshold $X_{u}$, when only bonds with coupling $J_{ij}$ less than
$X_{u}$ are considered. There is also a critical threshold $X_{u}^{c}$
that for $X_{u}<X_{u}^{c}$, the network is disconnected.

  In a typical random graph model, the degrees follow
a binomial distribution $B(k,p)$ \cite{feller1968anintroduction},
where $k$ is the degree and $p$ is the probability of having an
edge. One of the earliest random graph models is the Erd\H{o}s--Rényi
model \cite{erdhos1959onrandom}, where the graphs is characterized
by a parameter $p$ for the probability of finding an edge between
a given pair of vertices and there is no correlation between any edges.
For networks of a large size $N$, a sharp transition occurs at the
threshold $p=p_{c}=\log N/N$: The networks will almost surely be
connected for $p>p_{c}$ and almost surely be disconnected for $p<p_{c}$
\cite{erdhos1959onrandom}.

\begin{figure}
\begin{centering}
\includegraphics[width=10cm]{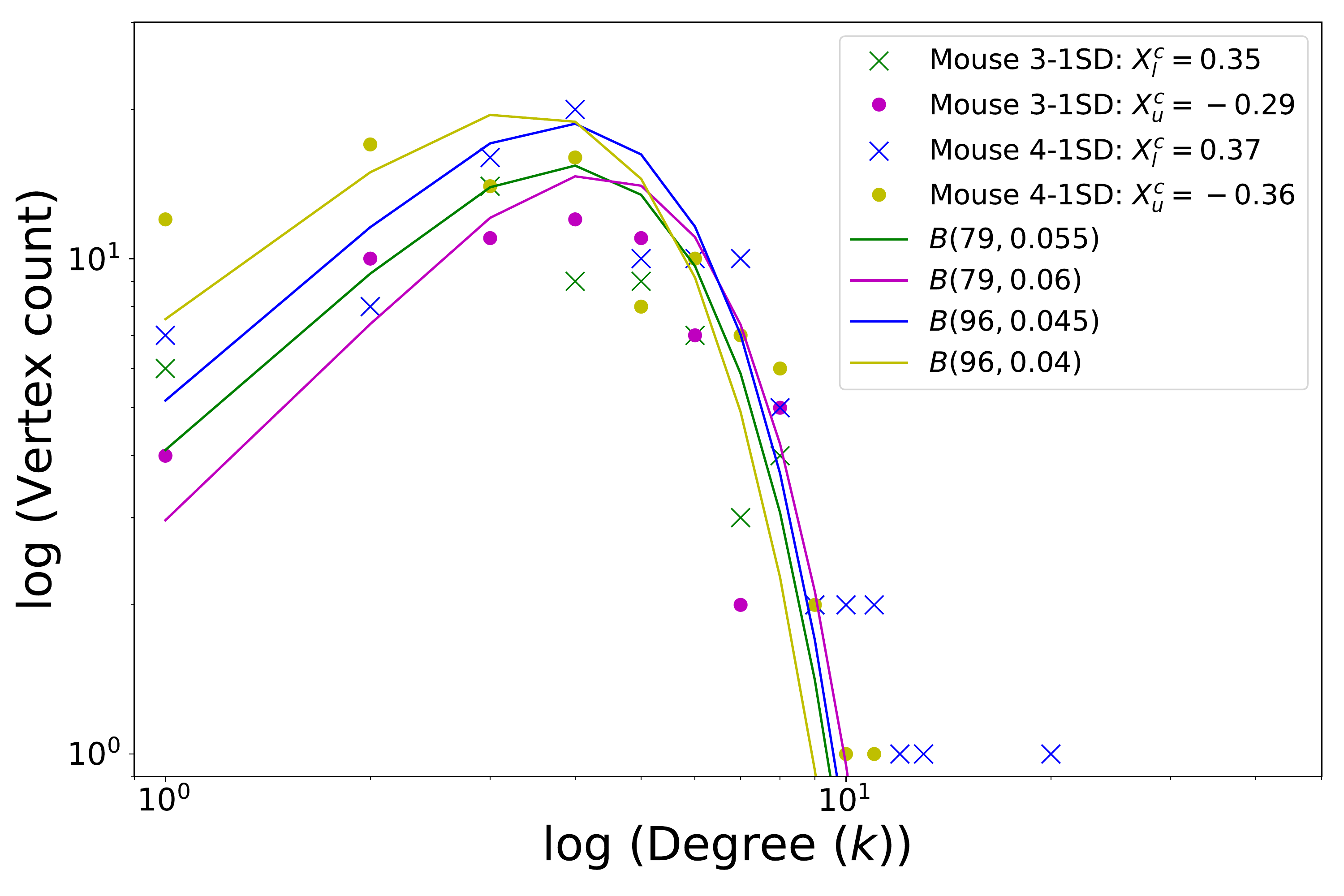}
\par\end{centering}
\caption{{\bf The log-log graph of degree distributions. }
The log-log plot of degree distributions at the critical threshold
$X_{l}^{c}$ in the excitatory networks and the critical threshold
$X_{u}^{c}$ in the inhibitory networks with their fitted binomial
distributions.}
\label{fig:ddd}
\end{figure}

In Fig.~\ref{fig:ddd}, the degree distributions at the critical thresholds
for all excitatory $X_{l}^{c}$ and inhibitory $X_{u}^{c}$ cases
are plotted and compared with the corresponding
binomial distribution.

\begin{figure}
\begin{centering}
\includegraphics[width=10cm]{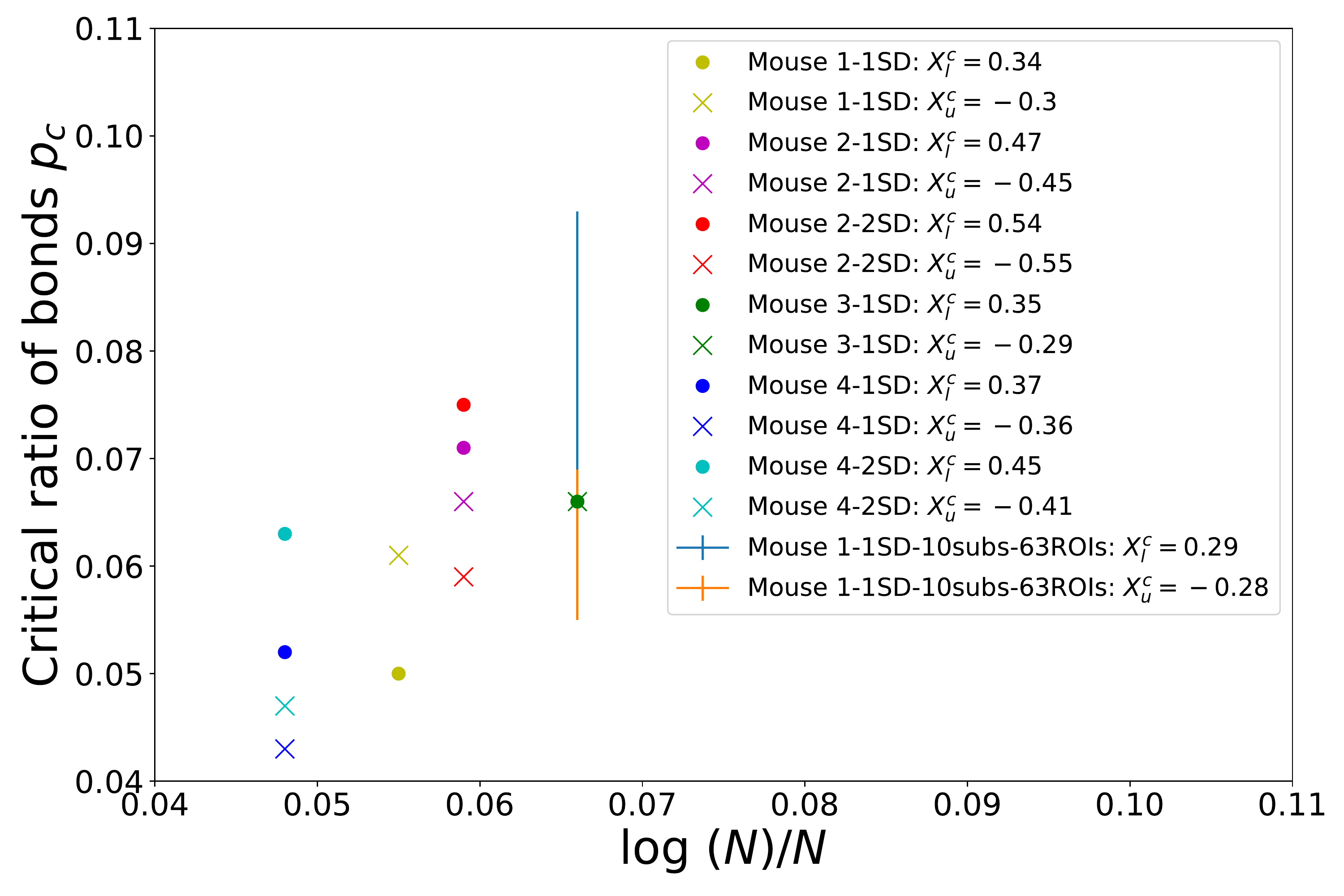}
\par\end{centering}
\caption{{\bf The prediction $\log(N)/N$ versus the critical fraction of bonds $p_{c}$. }
The critical fraction of bonds $p_{c}$ when the networks start to break
apart from one connected component versus the prediction $\log(N)/N$
from the random network model with $N$ being the system size.}
\label{fig:pN}
\end{figure}

  The fraction of bonds between all possible pairs of vertecis in the network is a function of 
$X_{l}$ ($X_{u}$) for the excitatory (inhibitory) network and it is considered the same as 
probability $p$ in the Erd\H{o}s--Rényi model. At $X_{l}^{c}$ and $X_{u}^{c}$, the fraction of bonds 
$p_{c}$ is the probability $p$ at the
critical thresholds. As seen in Fig.~\ref{fig:pN}, the values of
$p_{c}$ for all the excitatory and inhibitory cases of all our experiments
are close to the predicted values $\log N/N$ \cite{erdhos1959onrandom}
from the random network model. The deviation is probably due to the small size of our networks.

\section{Discussion\label{sec:Discussion}}

While criticality have been observed in neural signals of many studies of brains, it
is generally not very clear how universal these phenomena of critical
states are. In the process of mapping the brain dynamic to simple
statistical model, many details and much information is discarded.
Furthermore, even with the ever improving technology in experimental
studies of brains, the observed signals generally can only represent
a small part of the entire system. In the current study, we verify
the robustness of the proximity to a critical state in the CA1 recordings
from four free-moving mice. We use different thresholds and different
ways of subsampling the available ROIs into different sizes. The critical
temperatures as determined by the peak of specific heat as well as
the peak of magnetization slope are summarized in Fig.~\ref{fig:Critical-temperature-determined}
for all considered cases of thresholding and subsampling of the data
from the four mice. Majority of the peak temperature values are within
10\% of $T=1$ that is the temperature used to fit the experimental data. 
Thus the proximity to a critical state
seems to be very robust against variations in measurement and analysis.

  Along with the critical state, we also verify the structure of the
coupling network of the statistical model is robust under random subsampling
down to size 8. When comparing the distribution of the coupling strength
of subsamples with the original model, we find the KS statistic remains
small or comparable to the KS statistic between the original model
and a Gaussian for subsample sizes down to 30 (see Fig.~\ref{fig:Distributions-of-coupling}).
For subsample size smaller than 30, the KS statistic of a subsample
increases quickly. This suggests that a minimal number of ROIs of 30 in
experimental measurement may already have the main properties of the model
represented. To understand
the significance of the properties of the structure of the coupling
network uncovered by the statistical modeling, the best-fit model is perturbed
by independently changing the shape of the distribution and
its network structure (see Figs.~\ref{fig:Reshaped} and \ref{fig:randomized-network}).
While this proximity to a critical state seems to be very robust against
variations in measurement, random shuffling of the bond coupling strength or assigning 
a value according to a Gaussian distribution will move the system far 
away from the critical state of the model.

  We further find that the observed
networks are generally more balanced than random shuffles of the same
networks. While that can not fully account for the observed proximity
to a critical state, making a shuffled network more balanced does
help to bring it closer to a criticality. On the other hand, from
the application of different thresholding criteria, the connectivity
of the networks appears well described by the random network model.
This suggests that the connectivity may not be specific at the level
of average path length or degree distribution. Thus, the factors contributing
to the ubiquity of the critical state may be more subtle than one
would expect. A possible explanation is that the integration of sensory
information through layers of neurons that precede the CA1 can effectively
act as a renormalization process and filter out the irrelevant interactions
or correlations leaving the most critical degrees of freedoms to be
represented by the activities of the observed neurons at CA1.

Facing with the increasing volume of biological or social data, statistical
modeling is currently the only quantitative and nontrivial approach
that can be generally applied without any domain knowledge. Our results
show that the application of statistical modeling to in vivo recording
of brains can yield model properties that are robust to the arbitrariness
and randomness of sampling and segmentation. Specifically, the network
structure of the model is found to be crucial to the critical state.
It would be interesting to find the basic principles required to 
construct such a network. It is also interesting to find biological 
conditions for the neurons to be at  different critical states or non-critical at all.

\section{Methods}
\subsection{Experimental setup and data processing\label{sec:Experimental-setup-and}}

The male C57BL/6J mice (aged 8--12 weeks) were purchased from National
Laboratory Animal Center Inc. (Taipei, Taiwan) and maintained under
control conditions as follows: environmental temperature between $21$--$25$
$^{\circ}$C, a relative humidity $60\pm10\%$, a 12:12-h light:dark
cycle (light on from 08:00 to 20:00), and ad libitum access to food
and water. All experimental procedures were performed according to
China Medical University guidelines for the Care and Use of Experimental
Animals and approved by the Institutional Animal Care and Use Committee
of China Medical University.

To detect cytoplasmic free calcium activity in pyramidal regions of
interest, 0.5 $\mu$L pENN. AAV. CamKII. GCaMP6f. WPRE. SV40
(Addgene, \#100834, physical titer: $\geq1.5\times10^{13}$ vg/mL)
was injected into unilateral dorsal CA1 of the hippocampus (coordination:
caudal from bregma -1.94 mm, lateral from bregma -1.25 mm, ventral
form dura 1.35 mm) at 0.1 $\mu$L/min for 7min in Zoletil-anesthetized
mice. Two weeks after virus transduction, a Gradient-Index (GRIN)
lens (1.8 mm diameter, 4.39 mm length, Edmund Optics Inc.) was implanted
into dorsal CA1 (coordination: caudal from bregma $-1.94$ mm, lateral
from bregma $-1.25$ mm, ventral form dura 1.6 mm). For the relief
of pain and excessive innate immune responses to the GRIN lens implantation,
anesthetized mice were given additional 10 mg/kg carprofen (Sigma-Aldrich),
0.2 mg/kg dexamethasone (Sigma-Aldrich) subcutaneously, and 10 mg/kg
enrofloxacin (China Chemical \& Pharmaceutical Co., Ltd.) intraperitoneally.
After GRIN lens implantation, 0.4 mg/mL enrofloxacin were administrated
in the drinking water for 7--10 days to prevent bacterial infection.
Related surgery information was obtained from the UCLA miniscope website
(\url{http://miniscope.org/index.php/Surgery_Protocol}). After 3-week
recovery period, we used the integrated miniature endoscope (miniscope
V3) to check GCaMP+ regions of interest and then attached the baseplate
to the skull of the lightly anesthetized mice with acrylic cement.
After baseplate fixation, mice were ready to subsequent behavioral
trainings.

The $30\times30\times30$ cm\textsuperscript{3} white square box
surrounding with 2 distal visual cues on the contralateral walls was
used to examine locomotor activity of awake-behaving mice. Behavior
videos were recorded using Logitech C270 Webcam about 70 cm above
the arena. After the 3-day habituation, the novel environment-induced
hyperactivities were significantly decreased in mice. 
On the 4th consecutive
day, behavior and calcium imaging videos were recorded simultaneously
for 5-20 minutes in mice. All behavioral and dynamic calcium imaging
data were analyzed by the UCLA Miniscope software based on the constrained
nonnegative matrix factorization for microendoscopic data (CNMF-E) (the open-source MATLAB analysis package
was obtained from \url{https://github.com/daharoni/Miniscope_Analysis}).
The main data analyzed in the present study included mice positions,
the intensity of calcium imaging data (the temporal traces C), and
the deconvolution of calcium imaging data (calcium spikes).

In the current study, we consider the results from four separate experiments
with four different mice. In experiment 1 (Mouse 1), 79 ROIs are identified
from a 10 min recording. 72 ROI for 10 min. in experiment 2 (Mouse 2),
63 for 20 min. (Mouse 3) in experiment 3 and 96 for 10 min. (Mouse 4)
in experiment 4.

For the statistical modeling in Methods,
we convert the calcium signal trace $[\text{Ca}^{2+}]_{i}$ of each
ROI, indexed by $i$, into binary states $s_{i}=\pm1$ using the standard
deviation (SD) of the trace as a threshold $\theta_{i}=\sigma\left([\text{Ca}^{2+}]_{i}\right)$,
\begin{equation}
s_{i}=\begin{cases}
+1, & [\text{Ca}^{2+}]_{i}>\theta_{i}\\
-1, & \text{otherwise}
\end{cases}.
\end{equation}
The collection of image frames of the CA1 recording becomes an ensemble
of states for the system of $N=79$, $72$, $63$ and $96$ spins
in the four experiments respectively. Details about
transform experimental recordings to binary data can be found in the Supplementary S1 Transform experimental recordings to binary data
.

\subsection{Statistical modeling and multi-staged Boltzmann learning}

\subsubsection{All-to-all spin-glass model}

To model the distribution of states for the experimental ensembles,
we use an all-to-all pair-interaction spin-glass model similar to
earlier works by, e.g., Schneidmanet et al., 2006 \cite{schneidman2006weakpairwise}. The energy or
Hamiltonian of the spin-glass model is given by 
\begin{equation}
E\left(\mathbf{s}\right)=-\sum_{i}h_{i}s_{i}-\sum_{i<j}J_{ij}s_{i}s_{j}, \label{allspinmodel}
\end{equation}
where spin at site $i$, $s_{i}=\pm1$, $h_{i}$ is the local magnetic
field at site $i$, and $J_{ij}$ is the coupling strength between
two spins $i$ and $j$. The second summation is over all $N\left(N-1\right)/2$
possible spin-pairs in the all-to-all system. The probability for
the system to be observed at a given state $\mathbf{s}=\left\{ s_{i}\right\} $
is given by the Boltzmann distribution 
\begin{equation}
P\left(\mathbf{s}\right)=e^{-\beta E\left(\mathbf{s}\right)}/Z,
\end{equation}
where the normalization factor $Z$ is the partition function 
\begin{equation}
Z=\sum_{\mathbf{s}}e^{-\beta E\left(\mathbf{s}\right)},
\end{equation}
and $\beta=T^{-1}$ gives the inverse temperature of the system.
We set $T= \beta= 1$ for the fitting of experimental data.

The best-fitted values of parameters $h_{i}$ and $J_{ij}$ in the
model (\ref{allspinmodel}) are obtained using a Boltzmann learning
\cite{ackley1985alearning} algorithm, which minimizes the KL divergence
between the observed and model state distributions, 
\begin{equation}
D_{\text{KL}}=\sum_{\mathbf{s}}P_{\text{data}}\left(\mathbf{s}\right)\ln\frac{P_{\text{data}}\left(\mathbf{s}\right)}{P_{\text{model}}\left(\mathbf{s}\right)}.
\end{equation}
The gradients of the KL divergence in the $h_{i}$ and $J_{ij}$ parameter
space are given by 
\begin{align}
\frac{\partial D_{\text{KL}}}{\partial h_{i}} & =m_{i}^{\text{data}}-m_{i}^{\text{model}},\label{eq:h-gradient}\\
\frac{\partial D_{\text{KL}}}{\partial J_{ij}} & =C_{ij}^{\text{data}}-C_{ij}^{\text{model}},\label{eq:j-gradient}
\end{align}
where $m_{i}\equiv\left\langle s_{i}\right\rangle $ is the magnetization
of spin $i$ and $C_{ij}\equiv\left\langle s_{i}s_{j}\right\rangle $
is the correlation between spins $i$ and $j$. The angle brackets
indicate averages over observed state configurations of the experiment
system and over (Markov-Chain Monte Carlo, MCMC) simulation-generated
configurations for the model system.

Various optimization methods based on gradient descent can be applied
to find the minimum where the gradients (\ref{eq:h-gradient}) and
(\ref{eq:j-gradient}) vanish, leading to $m_{i}^{\text{data}}=m_{i}^{\text{model}}$
and $C_{ij}^{\text{data}}=C_{ij}^{\text{model}}$. However for sizable
systems, the convergence of the traditional Boltzmann learning can
be slow or unstable with a fixed learning rate. Also for each step
of the BL iteration, an extensive MCMC run can also be required for
reaching desirable accuracies of the magnetization and correlation
estimates of the model. We take a multi-staged Boltzmann learning
approach in this paper that uses shorter MC runs for earlier BL iterations
to save time at the cost of reduced accuracy and uses longer MC runs
at the later stages of the BL to guarantee the precision of the convergence.
Details are given in the the Supplementary S1 The multi-staged Boltzmann learning.

\section*{Acknowledgments}

We are grateful for the funding support by the Ministry of Science
and Technology of Taiwan (MOST) to TKL and CCC under the grant no. 108-2321-B-010-009-MY2,
to YLC under grand no. 109-2123-M-001-001, and to DCW under grand no. 107-2320-B-039-061-MY3.
Also, DCW is supported by the National Health Research Institutes under grant no. NHRI-EX110-10815NI
and in part by the China Medical University Hospital under grant no. DMR-108-102.

\section*{Supplementary information}
\section {Supplementary Discussion S1: Statistical modelling and multi-staged Boltzmann learning}

\subsection{Transform experimental recordings to binary data}

\begin{figure}[!htbp]
\begin{centering}
\includegraphics[width=12cm]{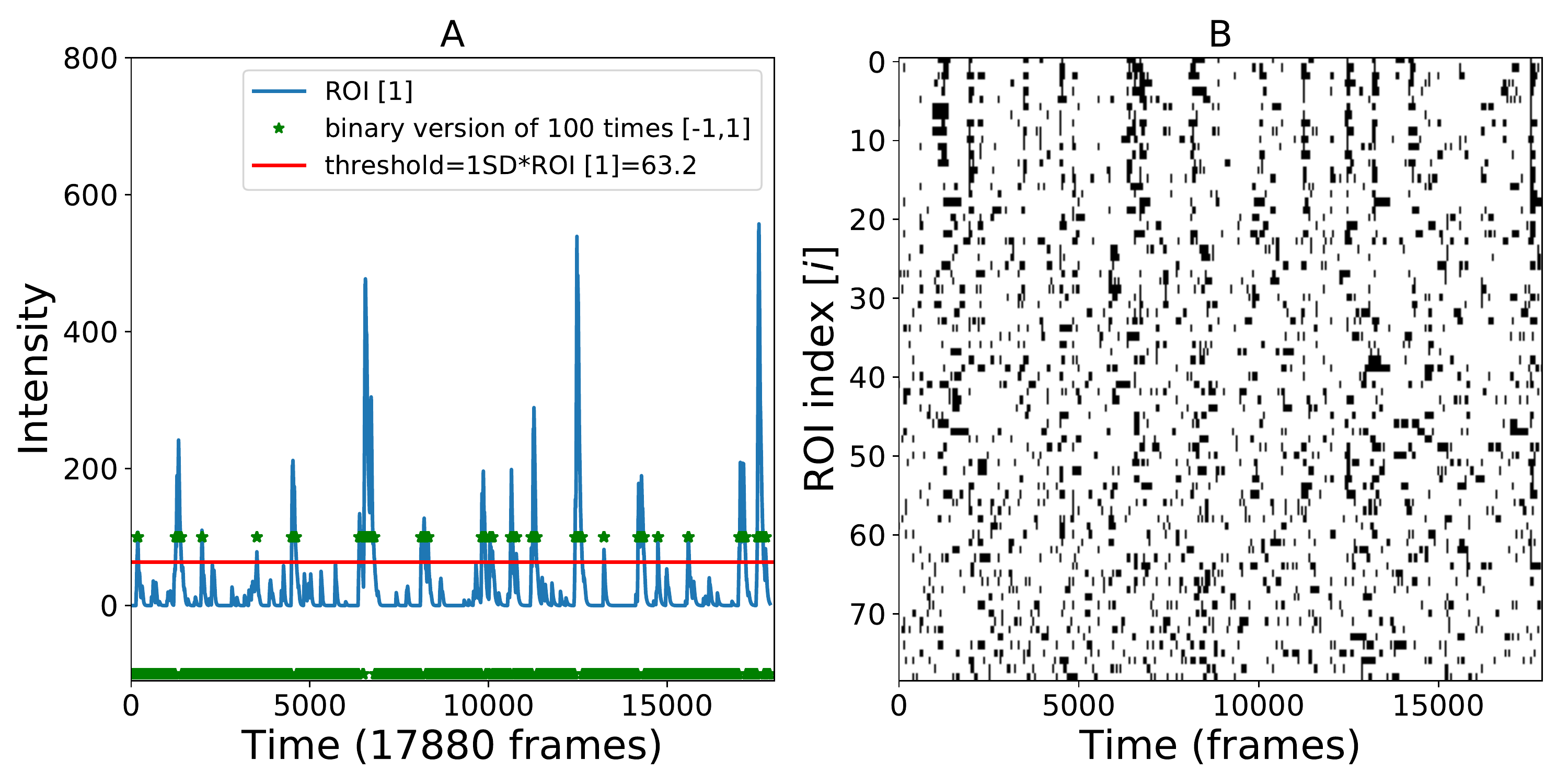}
\par\end{centering}
\caption{\textbf{A}. The calcium intensity of raw data of ROI{[}1{]} at 17880
time frames with the threshold= 1SD{*}ROI{[}1{]} at 63.3 (red line)
and its binary states (-1,1) times 100 as green points. \textbf{B}.
The binary version of 79 ROIs use a threshold to each ROI 1SD{*}ROI{[}$i${]},
$i=0,1,2,\ldots,78$ in 17880 time frames (10 minutes).\label{fig:binary}}
\end{figure}

We analyze data extracted from a genetically-encoded calcium ($\text{Ca}^{2+}$)
indicator, GCaNP6f and observe the dynamics of calcium in local CA1
regions of mice that are awake and freely moving in an open box. Unless
otherwise mentioned below, we present details of our analysis on the
Mouse 1 mouse data while similar results have also been obtained
from the other three data sets. For our analysis, the calcium signal
is first converted to binary data frames or states using the standard
deviation of each ROI (neuron) as a threshold. Supplementary Fig.~\ref{fig:binary}A
shows the results for ROI 1, with the blue line representing the raw
signal intensity of calcium imaging data at ROI 1 from a video of
17880 images frames (about 10 minutes with 30 frames per second).
The green stars show the binary states {[}-1, 1{]} scaled up by 100
intensity units representing the {[}active, silent{]} states, which
are obtained by applying the threshold of $\sigma\left([\text{Ca}^{2+}]_{1}\right)\approx63.2$
marked with the red line. Supplementary Fig.~\ref{fig:binary}B shows the binary
states $s_{i}$ of all ROIs (neurons) for all data frames with black
or white marking the active $s_{i}=+1$ or inactive $s_{i}=-1$ states
when the intensity for an ROI $i$ is respectively above or below
its threshold $\theta_{i}=\sigma\left([\text{Ca}^{2+}]_{i}\right)$.
We assign a vector of spin variables $\mathbf{s}=(s_{1},\ldots,s_{N})$
with $s_{i}=1$ for the $i$th active ROI and $s_{i}=-1$ for the
$i$th silent ROI in 79 ROIs. 

\subsection{The multi-staged Boltzmann learning}

In traditional Boltzmann learning (BL) algorithm, the model statistics
$m_{i}^{\text{model}}$ and $C_{ij}^{\text{model}}$ in the gradients
(Eqs.~(8) and (9) in the main text) are calculated using Markov-chain
Monte Carlo (MCMC) method with a fixed large number of steps to ensure
their precision. An iteration in the BL includes the updates 
\begin{eqnarray}
h_{i} & \leftarrow & h_{i}+\eta\left(m_{i}^{\text{data}}-m_{i}^{\text{model}}\right)\label{eq:bl_h}\\
J_{ij} & \leftarrow & J_{ij}+\eta\left(C_{ij}^{\text{data}}-C_{ij}^{\text{model}}\right)\label{eq:bl_j}
\end{eqnarray}
of the parameters of the model, which is repeated until the model
statistics match the data with desirable precision. To speed up this
process, instead of using a fixed large number of steps in MCMC throughout
all iterations of BL, we use a variable number of MCMC steps, adaptive
on the size of the difference between the model and the data statistics.
Specifically, we use the root-mean-square difference $d_{\text{rms}}\equiv C_{\text{rms}}+m_{\text{rms}}$,
with

\begin{eqnarray}
m_{\text{rms}} & \equiv & \sqrt{\frac{1}{N}\sum_{i}\left(m_{i}^{\text{data}}-m_{i}^{\text{model}}\right)^{2}}\label{eq:rmsM}\\
C_{\text{rms}} & \equiv & \sqrt{\frac{2}{N\left(N-1\right)}\sum_{i<j}\left(C_{ij}^{\text{data}}-C_{ij}^{\text{model}}\right)^{2}}\label{eq:rmsC}
\end{eqnarray}
to define the distance between the model and the data.
\begin{table}
\centering{}\caption{Variable number of MCMC steps in each BL iteration, adaptive to the
RMS error in BL \label{tab:Variable-number-of}}
\begin{tabular}{c|c}
$d_{\text{rms}}$ & Number of MCMC steps $T$\tabularnewline
\hline 
$0.05\leq d_{\text{rms}}$ & 10000\tabularnewline
$0.009\leq d_{\text{rms}}<0.05$ & 100000\tabularnewline
$0.005\leq d_{\text{rms}}<0.009$ & 1000000\tabularnewline
$0.004\leq d_{\text{rms}}<0.005$ & 2000000\tabularnewline
$0.003\leq d_{\text{rms}}<0.004$ & 3000000\tabularnewline
$0.002\leq d_{\text{rms}}<0.003$ & 4000000\tabularnewline
$d_{\text{rms}}<0.002$ & 5000000\tabularnewline
\end{tabular}
\end{table}

We start with $h_{i}=0$ and $J_{ij}=0$ as the initial condition
of the BL. For each iteration, depending on $d_{\text{rms}}$, the
MCMC runs for a variable number of steps $T$ as listed in Supplementary Table \ref{tab:Variable-number-of}
and the parameters $h_{i}$ and $J_{ij}$ are updated with rules of
Eqs.~\eqref{eq:bl_h} and \eqref{eq:bl_j} using $\eta=0.01$ as
the learning rate. After 100000 BL iterations, the set of $h_{i}$
and $J_{ij}$ that produced the minimum error $d_{\text{rms}}$ is
chosen as the best parameters of our model if this minimum $d_{\text{rms}}$
is less or equal to $0.003$. Otherwise, the set of $h_{i}$ and $J_{ij}$
is used as the initial condition for an extra BL run with learning
rate $\eta=0.003$ and a fixed number of MCMC steps $T=7000000$ in
each iteration until the error satisfies $d_{\text{rms}}<0.003$.
\begin{figure}
\begin{centering}
\includegraphics[width=9cm]{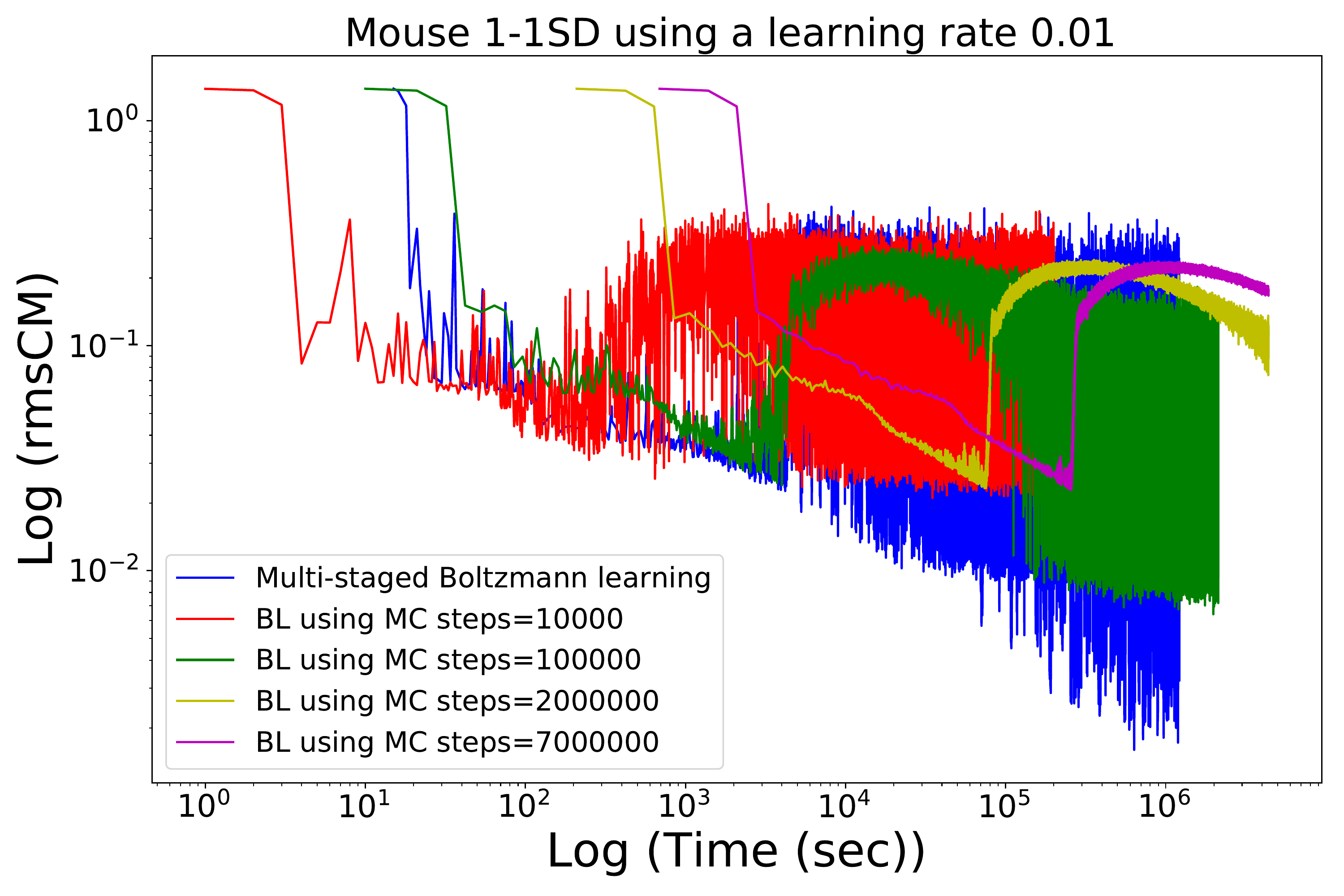}
\par\end{centering}
\caption{The log-log plot of the root mean-square difference versus the computer
running time (sec) using multi-staged BL (blue color) and other traditional
BL algorithm using different Monte Carlo (MC) steps. Traditional BL
algorithms seem to be easier to get trapped in local minima. Our multi-staged
MC method seems to be more efficient and also get consistent results
for different runs. \label{fig:Root-mean-square-difference}}
\end{figure}

The convergence of the model distribution to the data distribution
is shown in Supplementary Fig.~\ref{fig:Root-mean-square-difference} for different
learning algorithms. The hybrid BL in Supplementary Fig.~\ref{fig:Root-mean-square-difference}
(blue line) converges faster than the traditional BL algorithm using
the rate $\eta=0.01$ and $T=7000000$ (purple). In many cases of
using traditional BL approach with 10000 MC steps, the result cannot
converge in our running time (the red line in Supplementary Fig.~\ref{fig:Root-mean-square-difference}).
The main idea of the multi-staged BL algorithm is to correlate the
learning rate with the desired accuracy represented by $d_{\text{rms}}$.

\subsubsection{Model parameters}

After the convergence of the multi-staged BL method, the best-fitted
parameters of local field $h$ and coupling $J$ are first grouped
into histograms. Then the histograms are normalized to have the total
area of value 1 to become the distributions shown in Supplementary Figs.~\ref{fig:hist}A
and \ref{fig:hist}B for local field $h$ and coupling $J$, respectively.
The histograms have used the bin size determined by 
\begin{equation}
\frac{\text{maximum value of \ensuremath{h_{i}} (and \ensuremath{J_{ij}})- minimum value of \ensuremath{h_{i}} (and \ensuremath{J_{ij}})}}{\text{number of bins}},\label{h}
\end{equation}
where the number of bins is 50. The bin size of local fields is 0.11
and couplings is 0.08 in these cases.

\begin{figure}
\begin{centering}
\includegraphics[width=12cm]{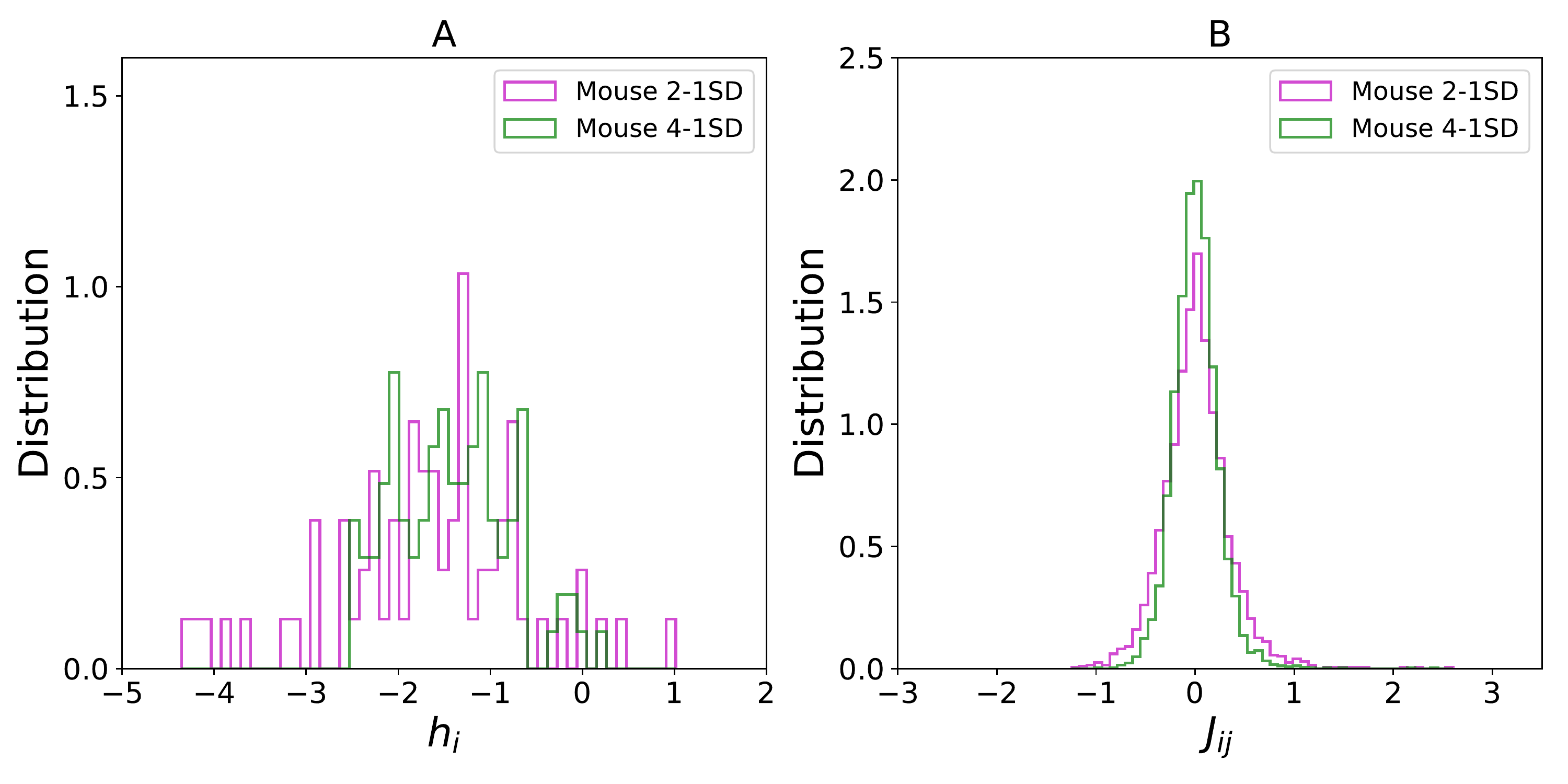}
\par\end{centering}
\caption{\textbf{A.} Distributions of the best-fitted local fields and \textbf{B.}
couplings. \label{fig:hist}}
\end{figure}

\section {Supplementary Discussion S2: Validation of the parameters}

\subsection{Matching the magnetization and covariance}

We use the all-to-all spin-glass model and the multi-staged Boltzmann
learning to get the best-fitted parameters by matching the observed
mean (magnetizations) and pairwise correlation with calculated mean
and pairwise correlation. The covariance is defined by 
\begin{equation}
\text{Cov}_{(X,Y)}=\frac{\sum_{i}^{N}(X_{i}-\overline{X})(Y_{i}-\overline{Y})}{N}
\end{equation}
,where $\overline{X}$ and $\overline{Y}$ are the means of $X$ and
$Y$, respectively. The correlation coefficient is $r=\frac{\text{Cov}_{(X,Y)}}{\sigma_{X}\sigma_{Y}}$,
where $\text{Cov}_{(X,Y)}$ is the covariance matrix and $\sigma_{X}$
and $\sigma_{Y}$ are the standard deviations of $X$ and $Y$, respectively.
The comparison of observed with the calculated magnetizations $M_{i}$
and covariance matrix $\text{Cov}_{ij}$ are shown in Supplementary Figs.~\ref{0425_3_MC}A
and \ref{0425_3_MC}B, respectively.
\begin{figure}
\begin{centering}
\includegraphics[width=12cm]{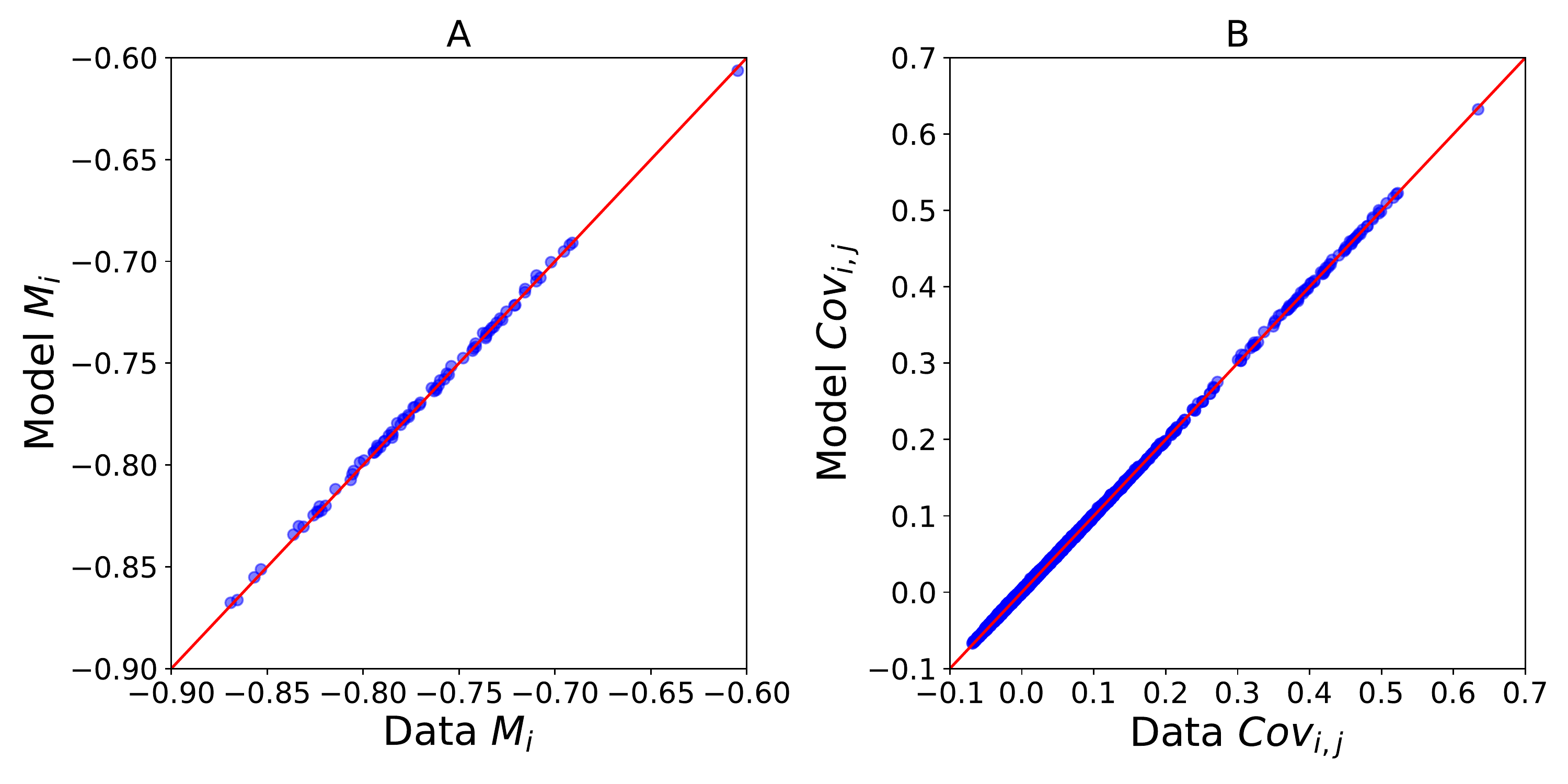}
\par\end{centering}
\caption{\textbf{A.} Matching the observed magnetizations with the calculated
magnetizations. \textbf{B.} Matching the observed covariance matrix
with the calculated covariance matrix using the data of Mouse 1-1SD.\label{0425_3_MC}}
\end{figure}

\subsection{Matching the high-order correlations}

For a more detailed comparison of the collective behaviors in data
with calculated populations, the triple correlation is calculated
for a triplet of spins as given by 
\begin{equation}
C_{ijk}=\left<\left(s_{i}-\left<s_{i}\right>\right)\left(s_{j}-\left<s_{j}\right>\right)\left(s_{k}-\left<s_{k}\right>\right)\right>.\label{triple}
\end{equation}
The result of comparing the observed with calculated triple correlations in the case of Mouse 1-1SD is shown in Supplementary Fig.~\ref{0425_3_34C} 
with correlation coefficient $r=0.84$. We included all possible triples of the observed data of 79 ROIs in 17880 frames and compared
with calculated states of 79 ROIs in 17880 frames.

\begin{figure}
\begin{centering}
\includegraphics[width=9cm]{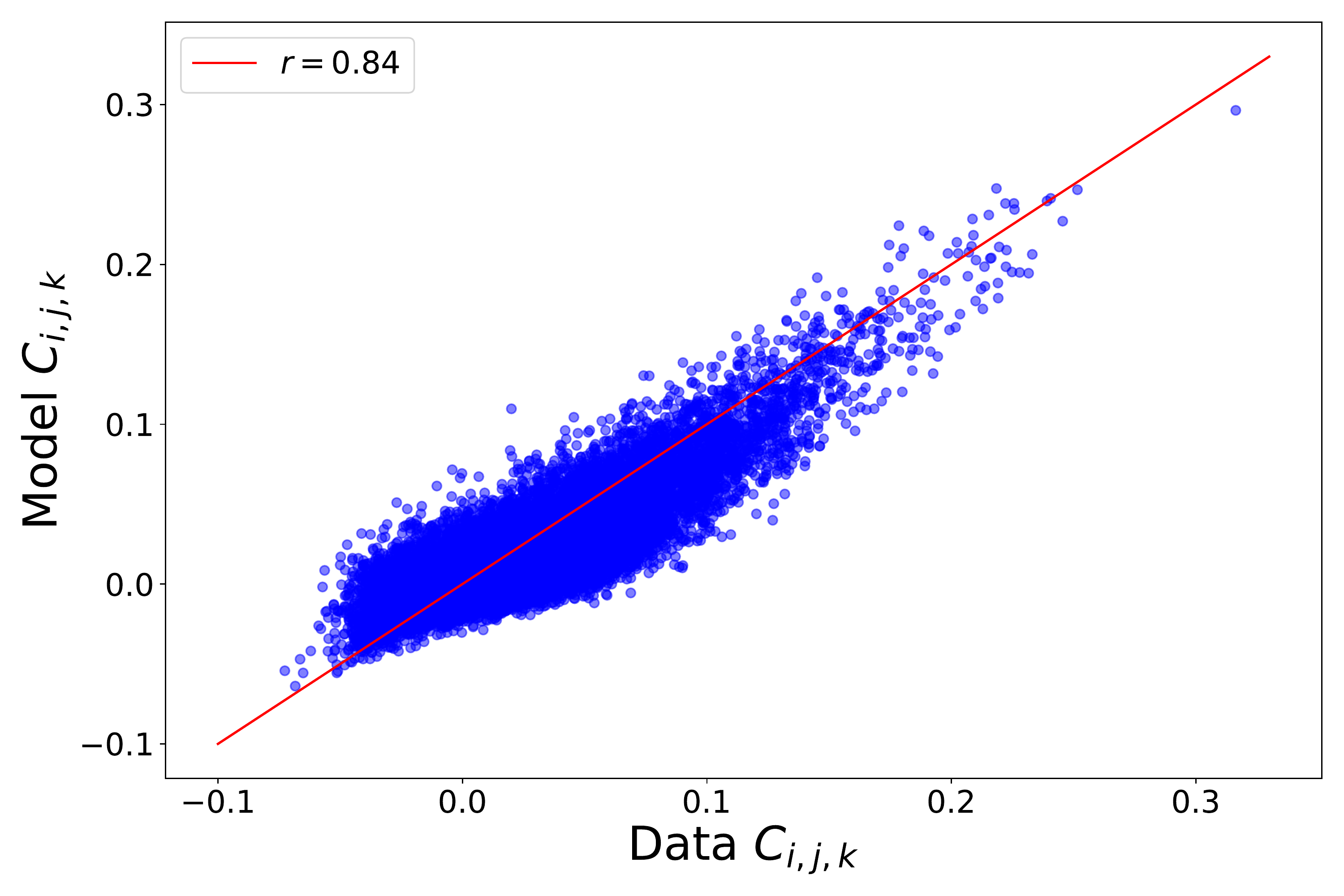}
\par\end{centering}
\caption{The observed versus calculated triple correlations using the data of Mouse 1-1SD.\label{0425_3_34C}}
\end{figure}

\section {Supplementary Discussion S3: Model predictions}

\subsection{$K$ simultaneously active neurons}

The probability of having $K$ out of $N$ ROIs active simultaneously
is shown in Supplementary Fig.~\ref{fig:0425_3K} for the data Mouse 1-1SD.
The agreement between model predictions and the data is reasonable
good until K=17. Following Meshulam et al., 2017~\cite{meshulam2017collective}, the uncertainty of the data
is estimated by using random half of the data. With 100 samples the
orange shade in Supplementary Fig.~\ref{fig:0425_3K} represents the one SD.

The probability of $K$ out of $N$ ROIs active simultaneously is
shown in Supplementary Fig.~\ref{fig:0425_3K} for the data Mouse 1-1SD.
Following Meshulam et al., 2017~\cite{meshulam2017collective}, the uncertainty
of the data is estimated by using random half of the data. With 100
samples the orange shade in Supplementary Supplementary Fig.~\ref{fig:0425_3K} represents the
one SD. The probability of having all ROIs in silent or $s_{i}=-1$
state for the model is $P(0)=0.021$ while the data has $0.015\pm0.001$.

\begin{figure}
\begin{centering}
\includegraphics[width=9cm]{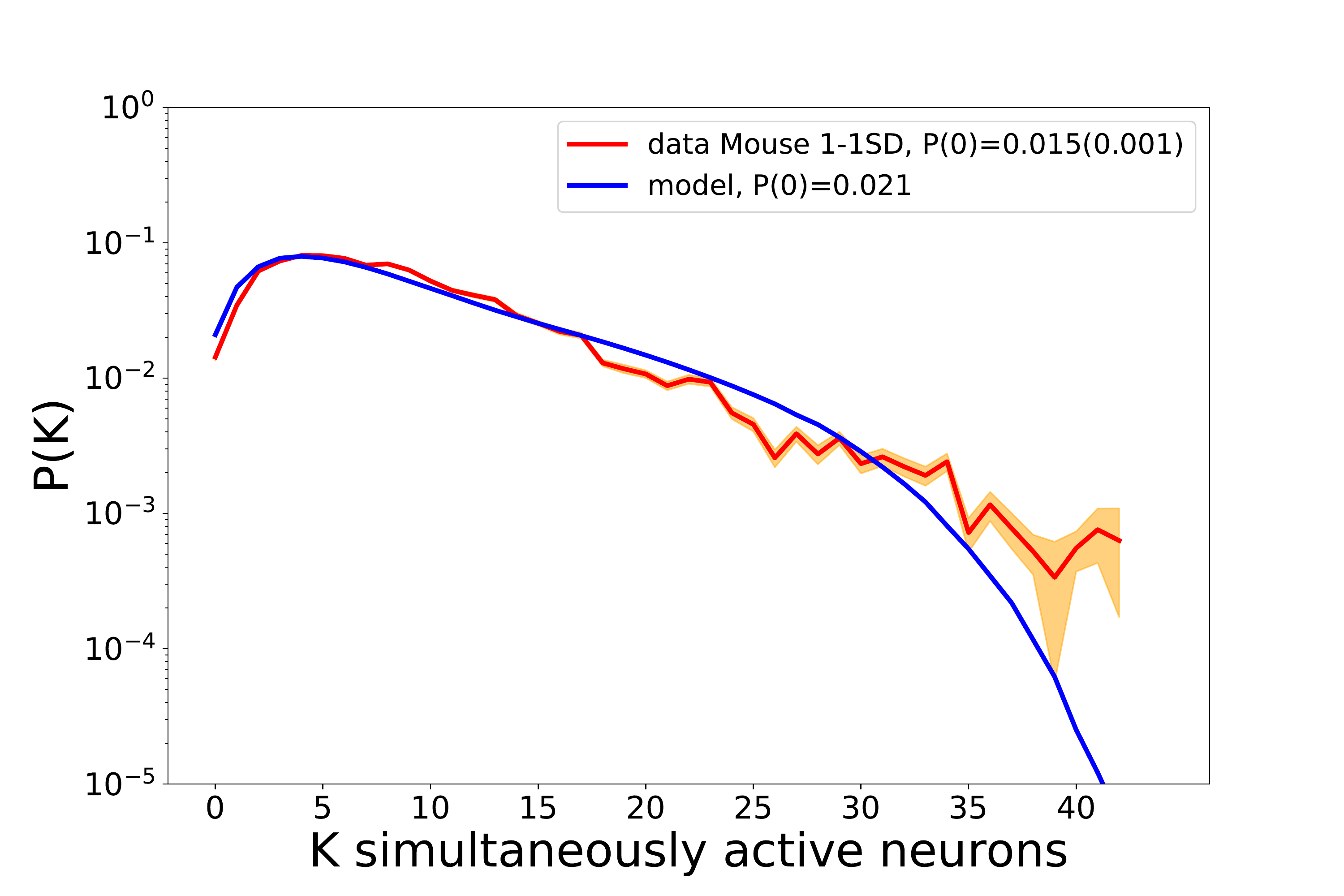}
\par\end{centering}
\caption{The probability of $K$ out of 79 neurons in the population are active
simultaneously of the model prediction (blue curve) and the mean of
data prediction (red curve) and the error bars of 1SD using random
halves of the data from 100 random trials (orange area). \label{fig:0425_3K}}
\end{figure}

\subsection{temperature dependence of the magnetization}

\begin{figure}
\begin{centering}
\includegraphics[width=9cm]{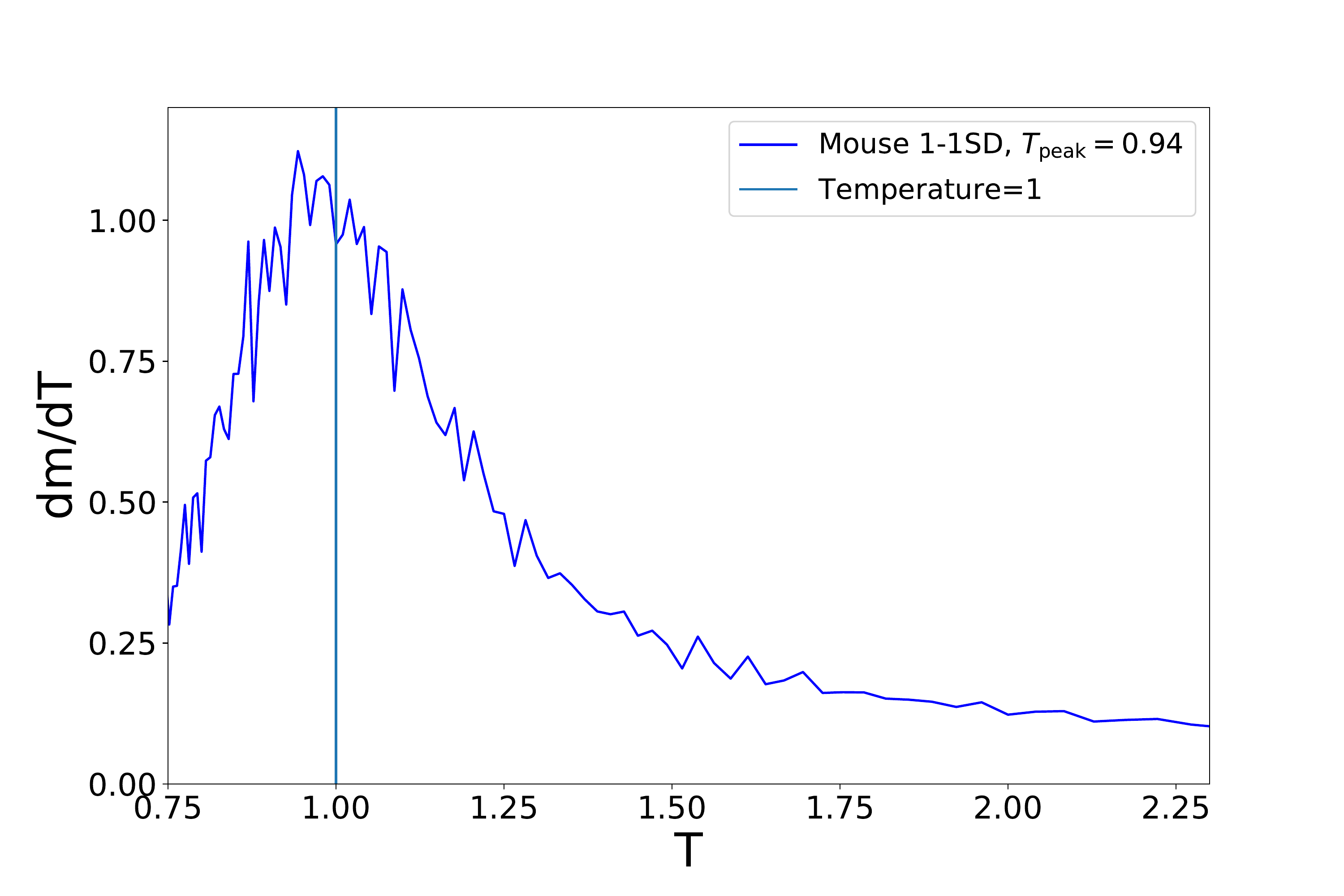}
\par\end{centering}
\caption{Magnetization slope $dm/dT$ versus temperature in the case Mouse 1-1SD.
\label{fig:0425_3dMdT}}
\end{figure}

The critical point of the spin glass system can be written as the
maximum slope $dm/dT$ (see, e.g., Huang, 1988 \cite{huang1988statistical}),
where the magnetization $m$ is the sum of all the spins, which is
equal to the observed mean activity of 79 ROIs in the case of Mouse 1-1SD.
Supplementary Fig.~\ref{fig:0425_3dMdT} shows the temperature dependence of magnetization
slope $dm/dT$. The maximum peak value is situated at $T_{\text{p}}=0.94$
which is very close to $T=1$ as marked by the vertical line. Note
that $T=1$ is the temperature the experimental data is fitted for.

\subsection{Temperature dependence of specific heat for subsamples}

The peak temperatures of the specific heat for subsamples of Mouse 4
data was referred to in Supplementary Supplementary Fig. 3 of the main text. Here we show the
detail temperature dependence of specific heat in Supplementary Supplementary Fig.~\ref{fig:Specific-heat-subsample-200303}.
Nine and ten subsamples with 79 and 63 ROIs are randomly selected
from the original data with 96 ROIs. Subsamples still show their proximity
to the critical state. The other three datasets also have similar
result.

\begin{figure}
\begin{centering}
\includegraphics[width=9cm]{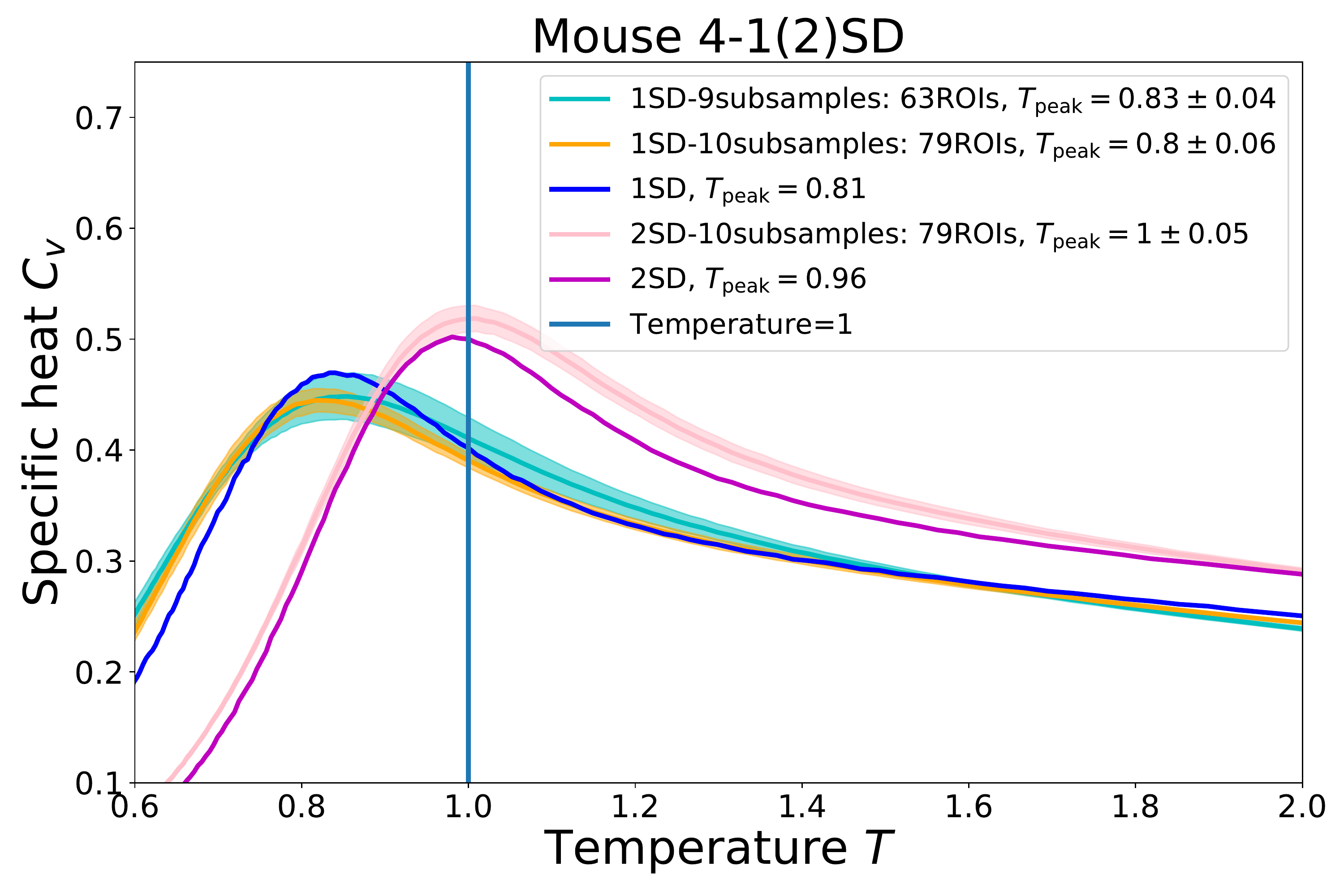}
\par\end{centering}
\caption{Specific heat curve from the statistical models for random subsamples
of measured ROIs in Mouse 4-1(2)SD experiment. The shaded regions
show the error bars of one standard deviation of uncertainty. \label{fig:Specific-heat-subsample-200303}}
\end{figure}

\section {Supplementary Discussion S4: Analysis of correlation of subsmaples with original data}

\subsection{Kolmogorov--Smirnov test}

In Sec.~V of the main text, the coupling distributions of the subsamples
and original data were compared with the Kolmogorov--Smirnov (KS)
test. Here we provide a little bit more information about the KS test
and its statistic. But readers are referred to, e.g., Sheskin, 2020 \cite{sheskin2020handbook}
for more details. The Kolmogorov--Smirnov statistic between two samples
of variable $x$ is defined as 
\begin{equation}
D_{n,m}=\mathrm{sup}_{x}\left|F_{1,n}(x)-F_{2,m}(x)\right|,
\end{equation}
where $F_{1,n}$ and $F_{2,m}$ are the cumulative distribution functions
of the two samples of size $n$ and $m$, respectively, and $\mathrm{sup}$
is the supremum function. It quantifies the difference between the
two distributions. When the measurements of each sample are drawn
independently from an underlying probability distribution, the KS
statistic can be related to the probability that the two samples are
drawn from the same distribution. In the current study, we are using
the KS statistic merely as a measure of how close two distributions
are to each other.

\subsection{Pearson correlation coefficients}

Pearson correlation coefficients between subsamples and original sample
are are presented in Fig. 6 of the main text. Here we provide a little
more detail of the calculations. Consider a subsample $\left\{ s'_{\alpha}\right\} $,
$\alpha=0,\ldots,N_{S}-1$ of an original data set $\left\{ s_{i}\right\} $,
$i=0,\ldots,N-1$ with the mapping $s'_{\alpha}=s_{i_{\alpha}}$ and
the condition $i_{\alpha}\neq i_{\beta}$ if $\alpha\neq\beta$. The
coupling strengths $J'_{\alpha\beta}$ of the statistical model for
the subsample are obtained through fitting the mean and covariance
of the size-$N_{S}$ spin-glass system to the statistics of the subset
of ROIs using the BL method. Since some spins of the original system
are not included, the coupling strength between two subsample spins
$\alpha$ and $\beta$ is generally different from the coupling between
the spins $i_{\alpha}$ and $i_{\beta}$ in the model of the original
size-$N$ system. To find how they are related, we calculate the Pearson
correlation coefficient between $J'_{\alpha\beta}$ and $J_{i_{\alpha}i_{\beta}}$
over all pairs $\alpha$,$\beta$ of the subsamples \cite{sheskin2020handbook},
\[
\text{\ensuremath{\rho=\frac{\sum_{\left\langle \alpha,\beta\right\rangle }\left(J'_{\alpha\beta}-\bar{J'}\right)\left(J_{i_{\alpha}i_{\beta}}-\bar{J}\right)}{\sqrt{\sum_{\left\langle \alpha,\beta\right\rangle }\left(J'_{\alpha\beta}-\bar{J'}\right)^{2}}\sqrt{\sum_{\left\langle \alpha,\beta\right\rangle }\left(J_{i_{\alpha}i_{\beta}}-\bar{J}\right)^{2}}}}}
\]
where $\bar{J'}\equiv2\sum_{\left\langle \alpha,\beta\right\rangle }J'_{\alpha\beta}/N_{S}\left(N_{S}-1\right)$,
and $\bar{J}\equiv2\sum_{\left\langle \alpha,\beta\right\rangle }J_{i_{\alpha}i_{\beta}}/N_{S}\left(N_{S}-1\right)$
are the mean coupling strength of subsample bonds in the subsample
and in the original system. For each subsample size $N_{S}$, the
calculation is repeated for 64 random subsamples and the mean and
variance of the Pearson correlation coefficients of their coupling
strengths with the original system are used to plot the line and the
shaded area in Fig.~6 of the main paper. The same calculations are
also performed for the local field $h_{i}$ with results shown in
Fig.~6 as well.

\subsection{Distributions of net coupling strength of ROIs}

The net coupling strength of ROIs in Mouse 1 data set was shown
in Fig. 9 of the main text. Below we include the results for the other
three datasets in Supplementary Fig.~\ref{fig:Net-ROI-coupling-strength-distri}.
\begin{center}
\begin{figure}
\begin{centering}
\includegraphics[width=12cm]{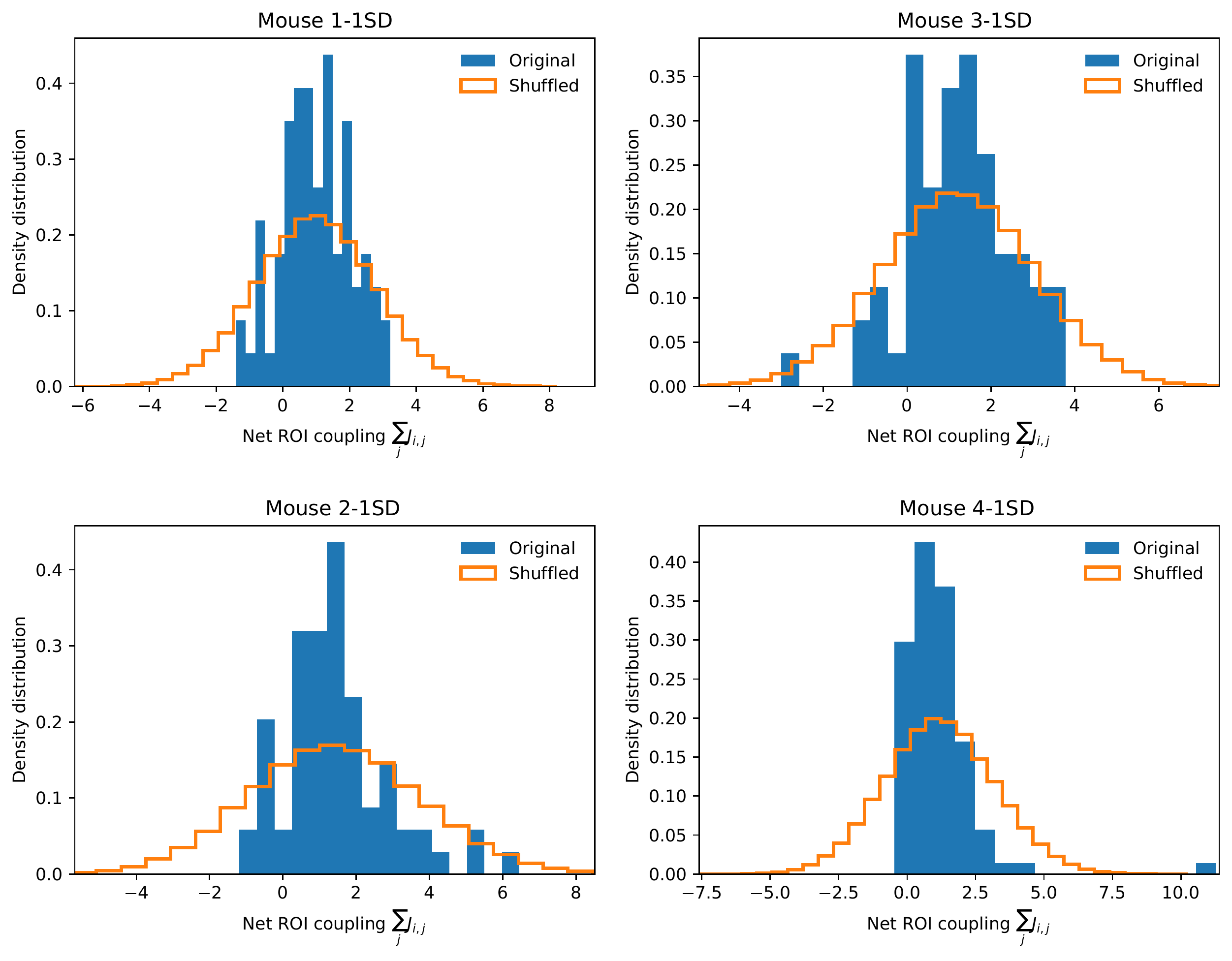}
\par\end{centering}
\caption{Net-ROI-coupling-strength distributions of the statistical model for
all the four datasets\label{fig:Net-ROI-coupling-strength-distri}}
\end{figure}
\par\end{center}

\section*{Author contributions statement}

YYM, NZ and DW carried out the mouse experiment and wrote Methods: Experimental setup and data processing.
YLC, CCC and TKL wrote the other part of the manuscript. All authors reviewed the manuscript. 
TKL conceived the idea and guided the project. 
YLC, and CCC performed the computations and all the figures in the main text and supplementary information.

\section*{Competing interests}
The authors declare no competing interests.

\bibliography{sample}

\end{document}